\documentclass[twocolumn,times]{aastex7}
\pdfoutput=1
\usepackage{amsmath}
\usepackage{graphicx}
\usepackage{indentfirst}
\usepackage{float}
\usepackage{graphicx}
\usepackage{subcaption}
\usepackage{caption}
\usepackage{multirow}
\usepackage{xcolor}
\usepackage{hyperref}

\usepackage{longtable}
\usepackage{booktabs}

\begin{document}

\title{Brightest GRB flare observed in GRB 221009A: bridge the last gap between flare and prompt emission in GRB}

\correspondingauthor{Shao-Lin Xiong, Shuang-Xi Yi}
\email{xiongsl@ihep.ac.cn, yisx2015@qfnu.edu.cn}

\author[0009-0002-6411-8422]{Zheng-Hang Yu}
\affil{State Key Laboratory of Particle Astrophysics, Institute of High Energy Physics, Chinese Academy of Sciences, 19B Yuquan Road, Beijing 100049, China}
\affil{University of Chinese Academy of Sciences, Chinese Academy of Sciences, Beijing 100049, China}
\email{}

\author[0009-0008-8053-2985]{Chen-Wei Wang}
\affil{State Key Laboratory of Particle Astrophysics, Institute of High Energy Physics, Chinese Academy of Sciences, 19B Yuquan Road, Beijing 100049, China}
\affil{University of Chinese Academy of Sciences, Chinese Academy of Sciences, Beijing 100049, China}
\email{}

\author[0000-0002-4771-7653]{Shao-Lin Xiong*} 
\affil{State Key Laboratory of Particle Astrophysics, Institute of High Energy Physics, Chinese Academy of Sciences, 19B Yuquan Road, Beijing 100049, China}
\email{}

\author{Shuang-Xi Yi*}
\affil{School of Physics and Physical Engineering, Qufu Normal University, Qufu, Shandong 273165, China}
\email{}

\author[0009-0008-6247-0645]{Wen-Long Zhang}
\affil{State Key Laboratory of Particle Astrophysics, Institute of High Energy Physics, Chinese Academy of Sciences, 19B Yuquan Road, Beijing 100049, China}
\affil{School of Physics and Physical Engineering, Qufu Normal University, Qufu, Shandong 273165, China}
\email{}

\author{Wen-Jun Tan}
\affil{State Key Laboratory of Particle Astrophysics, Institute of High Energy Physics, Chinese Academy of Sciences, 19B Yuquan Road, Beijing 100049, China}
\affil{University of Chinese Academy of Sciences, Chinese Academy of Sciences, Beijing 100049, China}
\email{}

\author[0000-0001-5348-7033]{Yan-Qiu Zhang}
\affil{State Key Laboratory of Particle Astrophysics, Institute of High Energy Physics, Chinese Academy of Sciences, 19B Yuquan Road, Beijing 100049, China}
\affil{University of Chinese Academy of Sciences, Chinese Academy of Sciences, Beijing 100049, China}
\email{}

\author[0009-0001-7226-2355]{Chao Zheng}
\affil{State Key Laboratory of Particle Astrophysics, Institute of High Energy Physics, Chinese Academy of Sciences, 19B Yuquan Road, Beijing 100049, China}
\affil{University of Chinese Academy of Sciences, Chinese Academy of Sciences, Beijing 100049, China}
\email{}

 \author[]{Hao-Xuan Guo}
 \affil{State Key Laboratory of Particle Astrophysics, Institute of High Energy Physics, Chinese Academy of Sciences, 19B Yuquan Road, Beijing 100049, China}
 \affil{Department of Nuclear Science and Technology, School of Energy and Power Engineering, Xi'an Jiaotong University, Xi'an, China}
 \email{}

 \author[0009-0004-1887-4686]{Jia-Cong Liu}
 \affil{State Key Laboratory of Particle Astrophysics, Institute of High Energy Physics, Chinese Academy of Sciences, 19B Yuquan Road, Beijing 100049, China}
 \affil{University of Chinese Academy of Sciences, Chinese Academy of Sciences, Beijing 100049, China}
 \email{}

 \author[]{Yang-Zhao Ren}
 \affil{State Key Laboratory of Particle Astrophysics, Institute of High Energy Physics, Chinese Academy of Sciences, 19B Yuquan Road, Beijing 100049, China}
 \affil{School of Physical Science and Technology, Southwest Jiaotong University, Chengdu 611756, China}
 \email{}

 \author[0009-0008-5068-3504]{Yue Wang}
 \affil{State Key Laboratory of Particle Astrophysics, Institute of High Energy Physics, Chinese Academy of Sciences, 19B Yuquan Road, Beijing 100049, China}
 \affil{University of Chinese Academy of Sciences, Chinese Academy of Sciences, Beijing 100049, China}
 \email{}

 \author[0000-0001-9217-7070]{Sheng-Lun Xie} 
 \affil{State Key Laboratory of Particle Astrophysics, Institute of High Energy Physics, Chinese Academy of Sciences, 19B Yuquan Road, Beijing 100049, China}
 \affil{Institute of Astrophysics, Central China Normal University, Wuhan 430079, China}
 \email{}

 \author[0000-0001-8664-5085]{Wang-Chen Xue}
 \affil{State Key Laboratory of Particle Astrophysics, Institute of High Energy Physics, Chinese Academy of Sciences, 19B Yuquan Road, Beijing 100049, China}
 \affil{University of Chinese Academy of Sciences, Chinese Academy of Sciences, Beijing 100049, China}
 \email{}

\author[]{Jin-Peng Zhang}
 \affil{State Key Laboratory of Particle Astrophysics, Institute of High Energy Physics, Chinese Academy of Sciences, 19B Yuquan Road, Beijing 100049, China}
 \affil{University of Chinese Academy of Sciences, Chinese Academy of Sciences, Beijing 100049, China}
 \email{}
 
 \author[0000-0002-8097-3616]{Peng Zhang}
 \affil{State Key Laboratory of Particle Astrophysics, Institute of High Energy Physics, Chinese Academy of Sciences, 19B Yuquan Road, Beijing 100049, China}
 \affil{College of Electronic and Information Engineering, Tongji University, Shanghai 201804, China}
 \email{}

\author[]{Zheng-Hua An}
\affil{State Key Laboratory of Particle Astrophysics, Institute of High Energy Physics, Chinese Academy of Sciences, 19B Yuquan Road, Beijing 100049, China}
 \email{}

\author{Ce Cai}
\affil{College of Physics and Hebei Key Laboratory of Photophysics Research and Application, 
Hebei Normal University, Shijiazhuang, Hebei 050024, China}
\email{}

\author[]{Pei-Yi Feng}
 \affil{State Key Laboratory of Particle Astrophysics, Institute of High Energy Physics, Chinese Academy of Sciences, 19B Yuquan Road, Beijing 100049, China}
 \affil{University of Chinese Academy of Sciences, Chinese Academy of Sciences, Beijing 100049, China}
 \email{}

\author[]{Min Gao}
\affil{State Key Laboratory of Particle Astrophysics, Institute of High Energy Physics, Chinese Academy of Sciences, 19B Yuquan Road, Beijing 100049, China}
 \email{}

\author[]{Ke Gong}
\affil{State Key Laboratory of Particle Astrophysics, Institute of High Energy Physics, Chinese Academy of Sciences, 19B Yuquan Road, Beijing 100049, China}
 \email{}

 \author[]{Dongya Guo}
\affil{State Key Laboratory of Particle Astrophysics, Institute of High Energy Physics, Chinese Academy of Sciences, 19B Yuquan Road, Beijing 100049, China}
 \email{}

\author[]{Yue Huang}
\affil{State Key Laboratory of Particle Astrophysics, Institute of High Energy Physics, Chinese Academy of Sciences, 19B Yuquan Road, Beijing 100049, China}
 \email{}

\author[]{Bing Li}
\affil{State Key Laboratory of Particle Astrophysics, Institute of High Energy Physics, Chinese Academy of Sciences, 19B Yuquan Road, Beijing 100049, China}
 \email{}
 
\author[]{Cheng-Kui Li}
\affil{State Key Laboratory of Particle Astrophysics, Institute of High Energy Physics, Chinese Academy of Sciences, 19B Yuquan Road, Beijing 100049, China}
 \email{}

\author[]{Xiao-Bo Li}
\affil{State Key Laboratory of Particle Astrophysics, Institute of High Energy Physics, Chinese Academy of Sciences, 19B Yuquan Road, Beijing 100049, China}
\email{}

\author[]{Xin-Qiao Li}
\affil{State Key Laboratory of Particle Astrophysics, Institute of High Energy Physics, Chinese Academy of Sciences, 19B Yuquan Road, Beijing 100049, China}
 \email{}

 \author[]{Ya-Qing Liu}
\affil{State Key Laboratory of Particle Astrophysics, Institute of High Energy Physics, Chinese Academy of Sciences, 19B Yuquan Road, Beijing 100049, China}
 \email{}

\author[]{Xiao-Jing Liu}
\affil{State Key Laboratory of Particle Astrophysics, Institute of High Energy Physics, Chinese Academy of Sciences, 19B Yuquan Road, Beijing 100049, China}
 \email{}

\author[]{Xiang Ma}
\affil{State Key Laboratory of Particle Astrophysics, Institute of High Energy Physics, Chinese Academy of Sciences, 19B Yuquan Road, Beijing 100049, China}
 \email{}

 \author[]{Wenxi Peng}
\affil{State Key Laboratory of Particle Astrophysics, Institute of High Energy Physics, Chinese Academy of Sciences, 19B Yuquan Road, Beijing 100049, China}
 \email{}

 \author[]{Rui Qiao}
\affil{State Key Laboratory of Particle Astrophysics, Institute of High Energy Physics, Chinese Academy of Sciences, 19B Yuquan Road, Beijing 100049, China}
 \email{}

\author[]{Li-Ming Song}
\affil{State Key Laboratory of Particle Astrophysics, Institute of High Energy Physics, Chinese Academy of Sciences, 19B Yuquan Road, Beijing 100049, China}
\affil{University of Chinese Academy of Sciences, Chinese Academy of Sciences, Beijing 100049, China}
\email{}

\author[]{Jin Wang}
\affil{State Key Laboratory of Particle Astrophysics, Institute of High Energy Physics, Chinese Academy of Sciences, 19B Yuquan Road, Beijing 100049, China}
 \email{}

 \author[]{Jin-Zhou Wang}
\affil{State Key Laboratory of Particle Astrophysics, Institute of High Energy Physics, Chinese Academy of Sciences, 19B Yuquan Road, Beijing 100049, China}
 \email{}

\author[]{Ping Wang}
\affil{State Key Laboratory of Particle Astrophysics, Institute of High Energy Physics, Chinese Academy of Sciences, 19B Yuquan Road, Beijing 100049, China}
 \email{}

\author[]{Xiang-Yang Wen}
\affil{State Key Laboratory of Particle Astrophysics, Institute of High Energy Physics, Chinese Academy of Sciences, 19B Yuquan Road, Beijing 100049, China}
\email{}

\author{Shuo Xiao}
\affil{School of Physics and Electronic Science, Guizhou Normal University, Guiyang 550001, China}
\affil{Guizhou Provincial Key Laboratory of Radio Astronomy and Data Processing, Guizhou Normal University, Guiyang 550001, China}
\email{}

\author[]{Sheng Yang}
\affil{State Key Laboratory of Particle Astrophysics, Institute of High Energy Physics, Chinese Academy of Sciences, 19B Yuquan Road, Beijing 100049, China}
 \email{}
 
\author[]{Shu-Xu Yi}
\affil{State Key Laboratory of Particle Astrophysics, Institute of High Energy Physics, Chinese Academy of Sciences, 19B Yuquan Road, Beijing 100049, China}
 \email{}

\author[]{Qi-Bin Yi}
\affil{State Key Laboratory of Particle Astrophysics, Institute of High Energy Physics, Chinese Academy of Sciences, 19B Yuquan Road, Beijing 100049, China}
\affil{School of Physics and Optoelectronics, Xiangtan University, Xiangtan 411105, China}
 \email{}

\author[]{Da-Li Zhang}
\affil{State Key Laboratory of Particle Astrophysics, Institute of High Energy Physics, Chinese Academy of Sciences, 19B Yuquan Road, Beijing 100049, China}
 \email{}

\author[]{Fan Zhang}
\affil{State Key Laboratory of Particle Astrophysics, Institute of High Energy Physics, Chinese Academy of Sciences, 19B Yuquan Road, Beijing 100049, China}
 \email{}

\author[]{Shuang-Nan Zhang}
\affil{State Key Laboratory of Particle Astrophysics, Institute of High Energy Physics, Chinese Academy of Sciences, 19B Yuquan Road, Beijing 100049, China}
\affil{University of Chinese Academy of Sciences, Chinese Academy of Sciences, Beijing 100049, China}
\email{}

\author[]{Yan-Ting Zhang}
 \affil{State Key Laboratory of Particle Astrophysics, Institute of High Energy Physics, Chinese Academy of Sciences, 19B Yuquan Road, Beijing 100049, China}
 \affil{University of Chinese Academy of Sciences, Chinese Academy of Sciences, Beijing 100049, China}
 \email{}
 
\author[]{Zhen Zhang}
\affil{State Key Laboratory of Particle Astrophysics, Institute of High Energy Physics, Chinese Academy of Sciences, 19B Yuquan Road, Beijing 100049, China}
 \email{}

\author[]{Xiao-Yun Zhao}
\affil{State Key Laboratory of Particle Astrophysics, Institute of High Energy Physics, Chinese Academy of Sciences, 19B Yuquan Road, Beijing 100049, China}
 \email{}

\author[]{Yi Zhao}
\affil{School of Computer and Information, Dezhou University, Dezhou 253023, Shandong, China}
 \email{}

\author[]{Shi-Jie Zheng}
\affil{State Key Laboratory of Particle Astrophysics, Institute of High Energy Physics, Chinese Academy of Sciences, 19B Yuquan Road, Beijing 100049, China}
 \email{}

\begin{abstract}

Flares are usually observed during the afterglow phase of Gamma-Ray Bursts (GRBs) in soft X-ray, optical and radio bands, but rarely in gamma-ray band. Despite the extraordinary brightness, GECAM-C has accurately measured both the bright prompt emission and flare emission of GRB 221009A without instrumental effects, offering a good opportunity to study the relation between them. In this work, we present a comprehensive analysis of flare emission of GRB 221009A, which is composed of a series of flares. Among them, we identify an exceptionally bright flare with a record-breaking isotropic energy $E_{\rm iso} = 1.82 \times 10^{53}$ erg of GRB flares. It exhibits the highest peak energy ever detected in GRB flares, $E_{\rm peak} \sim 300$ keV, making it a genuine gamma-ray flare. It also shows rapid rise and decay timescales, significantly shorter than those of typical X-ray flares observed in soft X-ray or optical band, but comparable to those observed in prompt emissions. Despite these exceptional properties, the flare shares several common properties with typical GRB flares. We note that this is the first observation of a GRB flare in the keV-MeV band with sufficiently high temporal resolution and high statistics, which bridges the last gap between prompt emission and flare.

\end{abstract}


\section{INTRODUCTION} 

Gamma-ray bursts (GRBs) are the most energetic and violent astrophysical phenomena in the universe. Within a few seconds, GRBs could release energy in the range $10^{50}$-$10^{54}$\,erg mostly in $\gamma$-ray band on cosmological distance scales \citep{A85_1992_binary,A86_2006_supernova,A83_2006_grb,a82_2013_swift_gbm,2018_grb_physics_zhang_bing,A81_2025_GRB,A202_Eiso}. GRBs radiation could be divided into two phases: prompt emission and afterglow emission. The prompt emission is produced within the jet by internal shocks resulting from collisions of relativistic shells \citep{A41_1994_grb_physics} or magnetic dissipation \citep{1_2011_ICMART}, and the multi-wavelength afterglow is radiated when the jet interacts with the interstellar medium and stellar wind, generating forward shocks and reverse shocks \citep{A87_1999_afterglow_shock,A88_2003_afterglow_shock,1_2013_afterglow,A89_2014_afterglow_shock,A90_2021_AFTERGLOW_shock,A91_2024_afterglow_shock,A92_2025_afterglow_shock}.

In addition to prompt emission and afterglow emission, there are flares observed during the afterglow phase in 0.2-10\,keV, whose origin is on debate. One leading interpretation attributes bright X-ray flares to strong internal shocks produced by late activities of the central engine, where collisions between relativistic shells generate intense emission \citep{A9_2005_fireball1,A10_2009_fireball2}. Alternatively, flares may originate from external shocks \citep{A106_external_shock}. The photospheric model has also been proposed \citep{A11_2016_photospheric} to interpret the flare. Spectral analyses of some X-ray flares have indeed revealed thermal component, supporting a photospheric contribution \citep{A107_termal1,A108_termal2}. Furthermore, magnetic dissipation mechanisms in the jet can also offer explanation of the flare \citep{A12_2006_magnetic}.

Observations from the X-Ray Telescope (XRT) onboard \textit{Neil Gehrels Swift Observatory} \citep{A93_Swift_2004} indicate that about half of the GRBs have X-ray flares \citep{A1_2006_flare_physics, 2018_grb_physics_zhang_bing}. These flares can occur at any time during the afterglow phase \citep{A2_2011_flare_every}, with one or multiple flares in a single GRB  \footnote{\url{https://www.swift.ac.uk/xrt_curves/}} \citep{A94_2007_Swift_LC}. 

These X-ray flares are characterized by a significant bump riding on the decaying afterglow. The light curves of X-ray flares show relatively rapid rises and sharp declines compared to the afterglow. \cite{C22_norris2} has shown that the rise index is generally greater than 3, while the decline index typically falls between 1 and 3. They occur predominantly between $10^2$ and $10^5$\,s after the initial prompt emission \citep{A3_2007_flare_statis1, A4_2007_flare_statis2}. The fluence of X-ray flares is generally around 10\% of the averaged fluence of the prompt emission phase \citep{A3_2007_flare_statis1, A4_2007_flare_statis2}. However, in some exceptional cases, the fluence of the flare can be comparable to that of the prompt emission, for example, the very luminous flare of GRB 050502B \citep{A5_2006_GRB050502B}.

Previous studies have found that the temporal behavior and spectral properties of X-ray flares exhibit similarities to those of the prompt emission \citep{A5_2006_GRB050502B,A3_2007_flare_statis1,A4_2007_flare_statis2}. For example, when fitted with the Norris05 function \citep{C21_norris1,C22_norris2}, the ratio of the rise time ($t_{\text{rise}}$) to the decay time ($t_{\text{decay}}$) for prompt emission is typically concentrated in the range of 0.4-0.5. Interestingly, this ratio for flares is  $t_{\text{rise}}/t_{\text{decay}} \sim 0.49$ \citep{C21_norris1,C22_norris2}, indicating comparable rise and decay timescales between the two phenomena. Additionally, systematic analysis of GRB catalogs show that the low-energy spectral index $\alpha$ of prompt emission is clustered around -1.1 \citep{A109_BATSE_catalog,A110_GBM_catalog1,A111_GBM_catalog2,A112_GBM_catalog3,A113_GBM_catalog4}. A similar clustering around -1 is also found in flare samples, further suggesting analogous spectral properties. \citep{A5_2006_GRB050502B,A3_2007_flare_statis1,A4_2007_flare_statis2,A14_2015_Mev_GeV}. Observational evidence also supports the similarity about consistent Hardness Ratio (HR) evolution pattern \citep{A1_2006_flare_physics, A5_2006_GRB050502B, A3_2007_flare_statis1, A4_2007_flare_statis2}. These results favor a common physical origin between X-ray flares and prompt emission. Comprehensive statistical analyzes of the physical parameters indicate that such flares most likely originate from late activities of the central engine \citep{A1_2006_flare_physics, A3_2007_flare_statis1, A4_2007_flare_statis2}. 

Studies in X-ray flare statistics show that the average flare luminosity follows a power-law decay with respect to the peak time \citep{A7_L_average}; both the rise and decay timescales exhibit positive linear correlations with the flare peak time \citep{A8_2016_Yi_flare_sample}.

In addition, flares could be detected in a wide range of wavelengths, spanning from radio, optical and soft X-ray to gamma-ray \citep{A1_2006_flare_physics,A25_optical_flare,A30_GeV_09A,A95_2022_optical_flare,A96_2012_radio_afterglow,A97_2025_radio_flare}. Some observations have revealed distinct flares in the soft $\gamma$-ray band with energies ranging from tens to hundreds of keV by \textit{Swift}/BAT observations \citep{A20_2024_gamma_ray_flare}. 
However, the statistics and temporal resolution of these observations are quite limited owing to the weakness of the flare. 
Moreover, until now, there has been no definitive evidence of prominent flares occurring specifically within the MeV-GeV energy band, while bright in the X-ray band. \citep{A14_2015_Mev_GeV}.

Although GRB flares and prompt emission share similarities in their temporal and spectral behaviors, several key distinctions remain.  First, flares typically occur between $10^2$ and $10^5$\,s after the burst onset \citep{A3_2007_flare_statis1,A4_2007_flare_statis2}, often after the afterglow component becomes dominant. In contrast, the prompt emission is observed predominantly at the beginning of the burst \citep{2018_grb_physics_zhang_bing}. Flares thus appear later and exhibit more isolated episodes of emission compared to the prompt component. Second, within the same burst, flares are generally softer in spectrum than the prompt emission, as indicated by softer spectral parameters such as fluence and $E_{\text{peak}}$ \citep{A5_2006_GRB050502B,A3_2007_flare_statis1,A4_2007_flare_statis2,C7_2014_falre_sample,A14_2015_Mev_GeV}. Third, flares often present a light curve structure superimposed on an underlying power-law decay of the afterglow component, in logarithmic space \citep{A1_2006_flare_physics,A3_2007_flare_statis1,A4_2007_flare_statis2}.

Despite many studies, there has been a lack of direct observational evidence to robustly demonstrate that flare is indeed a part of prompt emission originating from late activities of the central engine. Such evidence includes rapid variations in the light curve and high-precision time-resolved spectrum, that is, exactly resembling the temporal and spectral properties of prompt emission. However, flares are usually not bright enough to be detected with high temporal resolution and sufficiently high statistics.

Fortunately, GRB 221009A has a very bright flare emission. Indeed, this flare itself has more fluence than the second brightest GRBs (i.e. GRB 230307A) \citep{1_09A_Gecam,A99_2023_GRB230307A,A98_2024_GRB230307A,A100_2025_GRB230307A,A101_2025_GRB230307A,A102_2025_grb230307_NSR}. Moreover, GECAM-C accurately measured both the flare and prompt emission of this burst without instrumental effects \citep{1_09A_Gecam}. The high quality GECAM-C data of both the prompt emission and flare in GRB 221009A and the exceptionally high statistics of the observed photons of the flare offer us a unique opportunity to explore the nature and physics of the GRB flare.

During the flare phase of GRB 221009A, previous studies have conducted observational analyses in the keV–MeV energy range using various instruments \citep{1_09A_Gecam,A43_09A_GBM,A19_2023_true_flare}. For instance, \cite{A19_2023_true_flare} used \textit{Fermi}/GBM data to analyze this period, reporting transitions between a power-law model and a Band function before and after the flare. However, detectors such as \textit{Fermi}/GBM suffered from saturation effects during the brightest phase of the flare, preventing them from a complete analysis for the full time range.

Although previous study using GECAM-C data have been performed, they primarily relied on the bspec data, which have a fixed time bin width of 1\,s \citep{1_09A_Gecam}. In this work, we use the high time resolution EVT (event) data from GECAM-C/GRD01 to analyze the flare phase. This approach allows us to achieve a time binning as fine as 0.1\,s, enabling more precise temporal and spectral investigations of the flare.

In this paper, we conducted a comprehensive analysis of the flare of GRB 221009A using GECAM-C data in the keV-MeV energy band (6\,keV-6\,MeV). 
This paper is organized as follows: Section~\ref{section2} describes the GECAM observations and provides details on the data analysis process for the GRB 221009A flare; Section~\ref{section3} presents the results of the data analysis; and presents the discussion of these results.

\section{Observation and Analyses}\label{section2}

\subsection{GRB 221009A}
GRB 221009A is the Brightest Of All Time (BOAT) GRB ever detected \citep{1_09A_Gecam,A16_09A_Swift,A43_09A_GBM,B20_LHAASO_09A,A42_09A_insight,A103_2025_09A_LAT}. \textit{Fermi}/GBM firstly reported this event by realtime in-flight trigger \citep{A15_09A_Fermi,A43_09A_GBM}. \textit{Swift}/BAT provided the first accurate localization by observing the afterglow emission \citep{A16_09A_Swift}, eventually leading to the measurement of redshift of $z = 0.151$ \citep{A17_09A_redshift1, A18_09A_redshift2,B51_09A_INTEGRAL}. This precise localization enabled many multi-wavelength observations. Remarkably, thousands of very high energy (VHE) photons (above 100 GeV) were detected from this burst by LHAASO \citep{B20_LHAASO_09A, LHAASO_09A_2,A42_09A_soc}. Specificially, \textit{Insight}-Hard X-ray Modulation Telescope (\textit{Insight}-HXMT) was triggered by GRB 221009A during its routine ground search at 13:17:00.050 UTC on October 9, 2022, which is considered as the trigger time of GRB 221009A (denoted as $T_0$) in the current work \citep{1_09A_Gecam,A42_09A_insight}. In addition, GRB 221009A was also detected by many other detectors: Konus-\textit{WIND} \citep{B71_KW_GCN1,B72_KW_GCN2,B82_kw_09A}, \textit{AGILE} (MCAL and GRID) \citep{B73_AGILE_GCN1,B74_AGILE_GCN2}, \textit{INTEGRAL} (SPI-ACS) \citep{B75_INTEGRAL_GCN}, \textit{Solar Orbiter} (STIX) \citep{B76_STIX_GCN}, \textit{Spektr-RG} (ART-XC) \citep{B77_spektr_GCN}, \textit{GRBAlpha} \citep{B78_GRBALPHA_GCN}, \textit{SIRI-2} \citep{B80_SIPI_GCN,B79_SIRI2_09A}, \textit{BepiColombo} (MGNS) \citep{B81_mgns_GCN} and so on. Due to the extreme brightness of GRB 221009A, most instruments experienced instrumental effects such as data saturation.

GRB 221009A is exceptionally bright that most high energy telescopes (e.g. \textit{Fermi}/GBM, \textit{Insight}-HXMT)  suffered instrumental effects (e.g., data saturation, pulse pileup) during its main burst epoch \citep[e.g.][]{A15_09A_Fermi, A42_09A_insight}. However, thanks to the dedicated designs for extremely bright bursts, GECAM-C made the uniquely accurate and high-resolution measurements of the main burst, and found that this GRB has the largest $E_{\text{iso}} = 1.5 \times 10^{55}$ erg \citep{1_09A_Gecam}. 

Moreover, accurate and high resolution data of GECAM-C have led to many discoveries of GRB 221009A: Through joint data analysis from GECAM-C and \textit{Insight}-HXMT and \textit{Fermi}/GBM, \cite{B9_2024_09A_afterglow} examined the detailed temporal and spectral characteristics of the early afterglow (from $T_0+660$\,s to $T_0+1860$\,s, where $T_0$ is the \textit{Insight}-HXMT/HE trigger time.), and, for the first time, revealed the unique spectral structure (a combination of cutoff power-law and power-law) of the early afterglow across the keV–MeV band and an achromatic (from keV to TeV) jet break feature at $T_0$+1243\,s; \cite{A43_09A_TeV_KeV} found that there is a close relation between the keV-MeV prompt emission and the TeV emission, lending direct evidence to the continuous energy injection to the external shock. Especially, \cite{B10_2024_09A_line} performed a joint analysis of GECAM-C and \textit{Fermi}/GBM data of GRB 221009A and revealed a series of the Gaussian emission line evolving from 37\,MeV to 6\,Mev as a power-law function of time with power-law index of -1, which is an unprecedented feature found in GRBs.

\subsection{Observation}

GECAM (Gravitational Wave High-energy Electromagnetic Counterpart All-sky Monitor) is a constellation of gamma-ray monitors designed to search for high energy electromagnetic counterparts (EMs) associated with gravitational waves (GWs) and fast radio bursts (FRBs). In addition to its primary scientific goal, GECAM is also capable of monitoring and locating GRBs, soft gamma-ray repeaters (SGRs), and other gamma-ray transient sources.

As the third instrument in the series, GECAM-C (also called HEBS) \citep{B2_2023_HEBS}, was launched onboard the Space Advanced Technology demonstration satellite  (SATech-01) on 27 July 2022 \citep{B3_2024_HEBS_launch}. 

GECAM-C is equipped with a total of 12 gamma-ray detectors (GRDs) based on SiPM technology \citep{B4_2022_GRD} and 2 charged particle detectors (CPDs) made with redout plastic scintillators by SiPM array \citep{B5_2022_CPD}, with all detectors installed on two domes in opposite sides of the satellite. 10 of 12 GRDs have two electron readout channels with different detection energy bands, referred to as high gain (HG) and low gain (LG), while the other 2 GRDs only have one readout channel. Previous studies have shown that GECAM-C has good temporal and spectral performance \citep{B6_2024_ground_HEBs,B7_2023_cross_Gecam,B8_2024_calibration_GECAM}.

The specialized design of GECAM-C for very bright bursts, combined with its specific operational mode in the Sun-synchronous orbit (SSO) (only the GRD01 detector was turned on) make GECAM-C successfully and accurately measured the GRB 221009A \citep{1_09A_Gecam}. GECAM-C provided precise observation of the main emission, flare emission, and a part of early afterglow \citep{1_09A_Gecam}.

Most of the flare emission of GRB 221009A occurred between $T_0 + 350$\,s and $T_0 + 600$\,s \citep{1_09A_Gecam}, as shown in Figure~\ref{fig_flux_all} and \ref{fig_lc}. During this period, GECAM-C entered an area with a low background count rate, from $T_0 + 329$\,s to $T_0 + 678$\,s. Throughout the flare, all GRDs and CPDs of GECAM-C were operational, enabling the collection of EVT data from each GRD. Note that the high peak count rate of the flare also caused data saturation in some instruments \citep{A43_09A_GBM}. Due to the high count rate of GRB 221009A, there were some instrumental effects during the main emission, which caused inaccuracies in the dead time calculation of GECAM-C/GRD01 HG channel emerged. Although these issues did not distort the spectral shape, they resulted in a lower count rate in the HG channel. However, during the flare of GRB 221009A, GECAM-C/GRD01 did not exhibit these issues. GECAM-C/GRD05 showed data saturation at the flare peak \citep{1_09A_Gecam}. In contrast, GECAM-C/GRD01 did not experience data saturation, making it very suitable for accurate and detailed analysis. Consequently, this study uses the EVT data from GECAM-C/GRD01 for the analysis of the flare \citep{1_09A_Gecam}.

During the detection of GRB 221009A, the SATech-01 satellite had been implementing a pointed observation, and GECAM-C remained almost a constant pointing direction during both the burst and the revisit orbit periods. Therefore, this allows us to use the data from the revisit orbit to estimate the background of GRB 221009A \citep{B9_2024_09A_afterglow}. In previous work, the revisit orbit method for background estimation has been proven to effectively measure background during the GRB 221009A \citep{1_09A_Gecam,B9_2024_09A_afterglow,B10_2024_09A_line}.

\subsection{Temporal analysis}\label{section2.2}

We used data from GECAM-C/GRD01 during the flare phase to analyze the light curve. 
As shown below (see {Figure~\ref{fig_flux_all}), these flares have exceptional brightness for GRB flares; therefore, we designate the flares of BOAT GRB 221009A as the BOAT flare. 
Analysis of the BOAT flare shows that the time interval of very high count rate is mainly concentrated between $T_0 + 500$\,s and $T_0 + 520$\,s.

The entire evolution of the BOAT flare is well measured within the main observation energy range of GECAM-C (6\,keV to 6\,MeV). We plotted the background-subtracted (net) light curve of the BOAT flare in a time bin width of 0.1\,s, revealing the temporal structure of the flare.

When displaying the light curves in different energy bands, the energy bands above MeV were plotted separately to display the light curve structure of the flare in the MeV energy band.  In this work, photons in the 100\,keV to 6\, MeV band are defined as soft $\gamma$-ray photons, while those in the 6\,keV to 100\,keV band are defined as hard X-ray photons. The ratio of soft $\gamma$-ray to hard X-ray photons in each time interval is defined as the spectral hardness. The hardness ratio was calculated to characterize the temporal evolution of the spectral hardness of the flare.

\begin{figure*}
  \centering
  \includegraphics[width=0.8\textwidth]{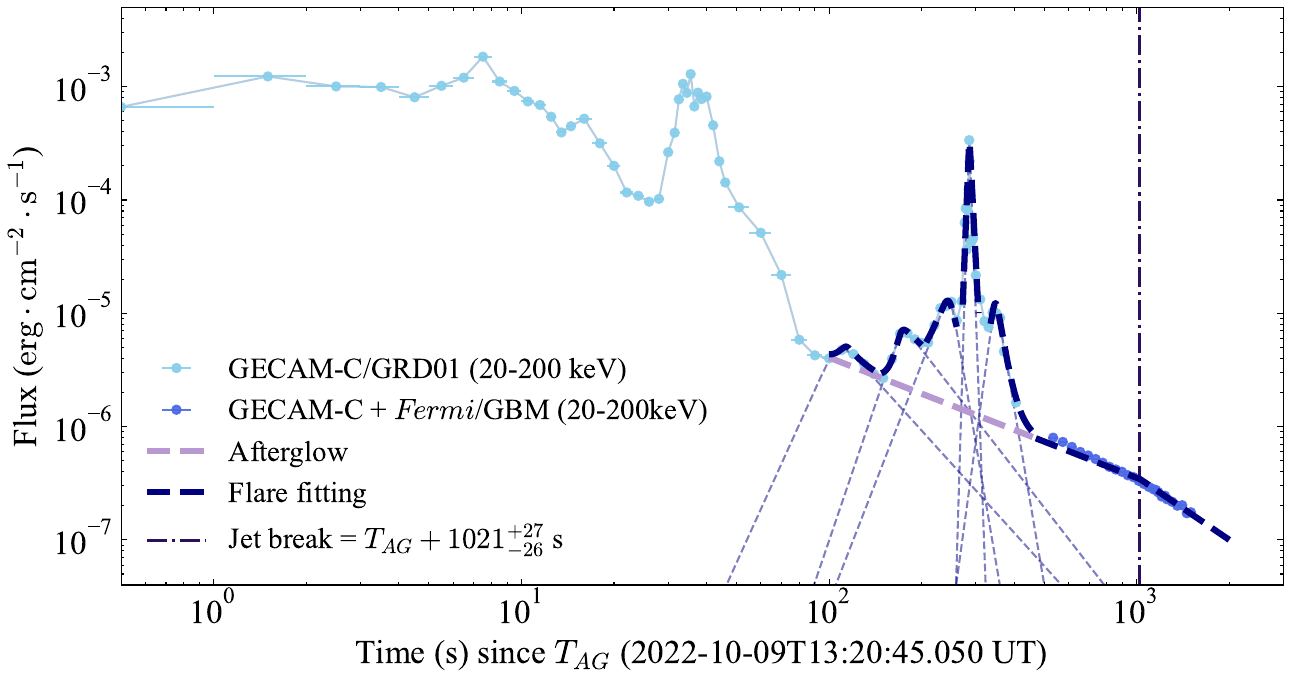}
  \caption{\label{fig_flux_all}\small Light curve of GRB 221009A from prompt emission to afterglow. The light blue line shows the GECAM-C/GRD01 flux (20-200\,keV) from $T_{\rm AG}$ to $T_{\rm AG}+415$\,s (data from \cite{1_09A_Gecam}). The dark blue dotted line fits entire flares and afterglow phase. The blue data points denote the flux (20-200\,keV) during the afterglow phase obtained from joint observations by GECAM-C and $Fermi$/GBM from $T_{\rm AG}+435$\,s to $T_{\rm AG}+1635$\,s. The light purple dotted line extends afterglow into the flare time period (data from \cite{B9_2024_09A_afterglow} ). Dark Blue vertical line marks the jet break time $T_{\rm AG} + 1021^{+27}_{-26}$\,s \citep{B9_2024_09A_afterglow}.}
\end{figure*}

\subsection{Spectral analysis}\label{2.3}

\subsubsection{Spectral models}

To fit the spectrum of the BOAT flare, we employed the Band function \citep{Band_function}, the cutoff power-law model (CPL) \citep{CPL_model}, the power-law model (PL) and the Blackbody (expressed as bbodyrad; BB) \citep{A201_BB_model}. The latest version of the GECAM-C CALDB and GECAMTools data analysis tools was used to generate the response matrix and spectrum files. Then, we used Xspec (V12.15.0) \citep{C5_1996_xspec} and PyXspec (V2.1.4) to fit the spectra of the flare. 

The Band, CPL, PL and BB are represented by Equations \ref{eq1}, \ref{eq2}, \ref{equ:pl_Model} and \ref{equ:bb_Model}, respectively:

The Band model is expressed as:

\begin{equation}
        N(E) =\begin{cases}
        A\left(\frac{E}{E_{\text{piv}}}\right)^{\alpha} \exp \left(-\frac{E}{E_{\text{c}}}\right), & E \leq(\alpha - \beta) E_{\text{c}}\\
        A\left[\frac{(\alpha - \beta) E_{\text{c}}}{E_{\text{piv}}}\right]^{\alpha - \beta} \exp (\beta - \alpha)\\\,\,\,\,\,\left(\frac{E}{E_{\text{piv}}}\right)^{\beta}, & E>(\alpha - \beta) E_{\text{c}} 
        \end{cases}\\\label{eq1}
\end{equation}

where $A$ is the normalization amplitude ($\rm photons \cdot cm^{-2} \cdot s^{-1} \cdot keV^{-1}$), $\alpha$ is the power law index \text{low-energy} and $\beta$ is the power law index \text{high-energy}, $E_{\rm c}$ is the characteristic energy in keV, $E_{\rm piv}$ is the pivot energy in keV and usually the value is 100 keV. $E_{\rm peak}$ =(2+$\alpha$)$E_{\rm c}$

The CPL model is expressed as:

\begin{equation}
N(E) = A \left( \frac{E}{E_{\text{piv}}} \right)^\alpha \exp \left( -\frac{E}{E_{\text{c}}} \right) \label{eq2}
\end{equation}

where $A$ is the normalization amplitude ($\rm photons \cdot cm^{-2} \cdot s^{-1} \cdot keV^{-1}$); $\alpha$ is the power law photon index; $E_{\rm c}$ is the characteristic energy in keV; $E_{\rm piv}$ is 100 keV.

The PL model is expressed as:

\begin{equation}  
N(E) = A \left( \frac{E}{E_{\text{piv}}} \right)^\alpha
\label{equ:pl_Model}
\end{equation}

where $A$ is the normalization amplitude ($\rm photons \cdot cm^{-2} \cdot s^{-1} \cdot keV^{-1}$); $\alpha$ is the power law photon index; $E_{\rm piv}$ is 100 keV.

The bbodyrad model is expressed as:

\begin{equation}   
N(E)=\frac{1.0344 \times 10^{-3}\times AE^2}{{\rm exp}(\frac{E}{kT})-1},
\label{equ:bb_Model}
\end{equation}

where $A=R^2_{\rm km}/D^2_{10}$ is the normalization constant, $R_{\rm km}$ is the source radius in km and $D_{10}$ is the distance to the source in units of 10 kpc, and $kT$ is the temperature in keV.

\subsubsection{Spectral fitting strategy}

In this analysis, we conducted an exhaustive fitting of the spectrum for the BOAT flare using various models across different time periods. 
To provide a comprehensive understanding of the characteristics of the spectrum, 
we set the time intervals for time-resolved spectroscopy using three kinds of time binning, denoted as S-1, S-2, and S-3 respectively, which is summarized in Table~\ref{table_spec}.

Since the observed data from the revisit orbit served as the background which follows a Poisson distribution, the \texttt{cstat} statistic is used for the spectral fitting \citep{1_09A_Gecam,B10_2024_09A_line}.

For three kinds of time bin S-1, S-2, S-3, we fit the spectrum for each time interval using seven models: Band, CPL, PL, Band+PL, CPL+PL, Band+BB and CPL+BB. The Bayesian information criterion (BIC) was computed for each model, which is defined as: $\text{BIC} = -2 \ln L + k \ln N$, where $L$ is the likelihood function, $k$ represents the number of free parameters in the model, and $N$ is the number of data points. The one with the minimum BIC value was selected as the best-fit model for the corresponding interval. The spectral fitting and the corresponding analysis results will be thoroughly discussed in Section~\ref{section3}.

\section{Results and discussion}\label{section3}

\subsection{A genuine flare}

\cite{1_09A_Gecam} preliminarily suggested that the emission from $T_0 + 350$\,s to $T_0 + 600$\,s is flare phase as it is much weaker than prompt emission and rides on the decaying afterglow. In another study \citep{A19_2023_true_flare} spectral analysis was implemented before and after the flares, and the spectral evolution exhibits a gradual transition between the PL model and the Band function. Prior to the burst, the spectrum is well fitted by a PL model over a certain period (from $T_0 + 320$\,s to $T_0 + 328$\,s,  where $T_0$ is the \textit{Fermi}/GBM trigger time), with spectral parameters consistent within errors with those of the post burst afterglow. This indicates that an afterglow component already dominated the emission for some time before the burst occurred. The flux of the power-law component before and after the burst can be well fitted by a power-law decay function, with only a few time intervals dominated by the power-law component in the Band+PL model. 

The evolution of the spectral fitting model, combined with the power-law decay of the afterglow light curves before and after the flare indicates that GRB 221009A had already entered the afterglow phase prior to the flare onset. This demonstrates a transition from the afterglow-dominant to the flare-dominant and then back to the afterglow-dominant.

\begin{figure*}
  \centering
\begin{tabular}{cc}
  \begin{minipage}{0.46\textwidth}
      \includegraphics[width = \textwidth]{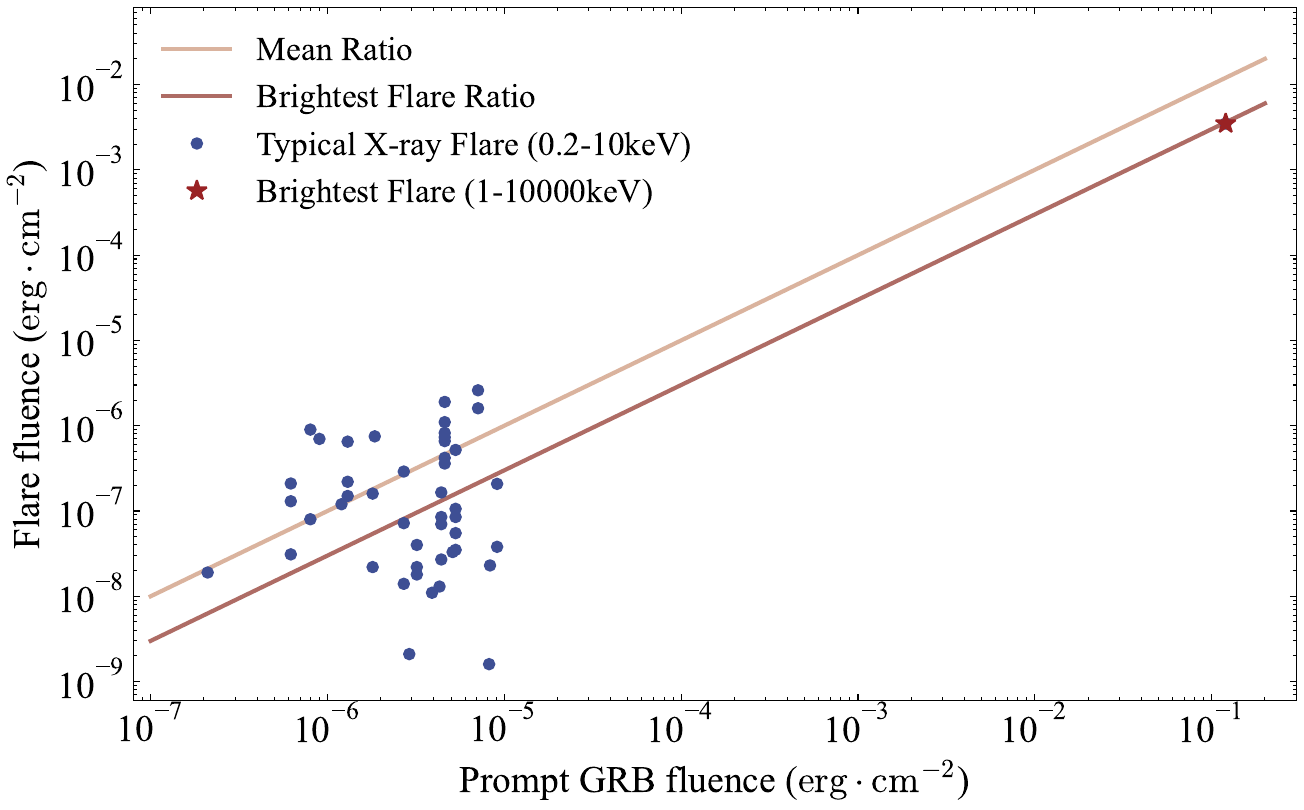}
      \subcaption{}
   \end{minipage}
    \begin{minipage}{0.44\textwidth}
        \includegraphics[width = \textwidth]{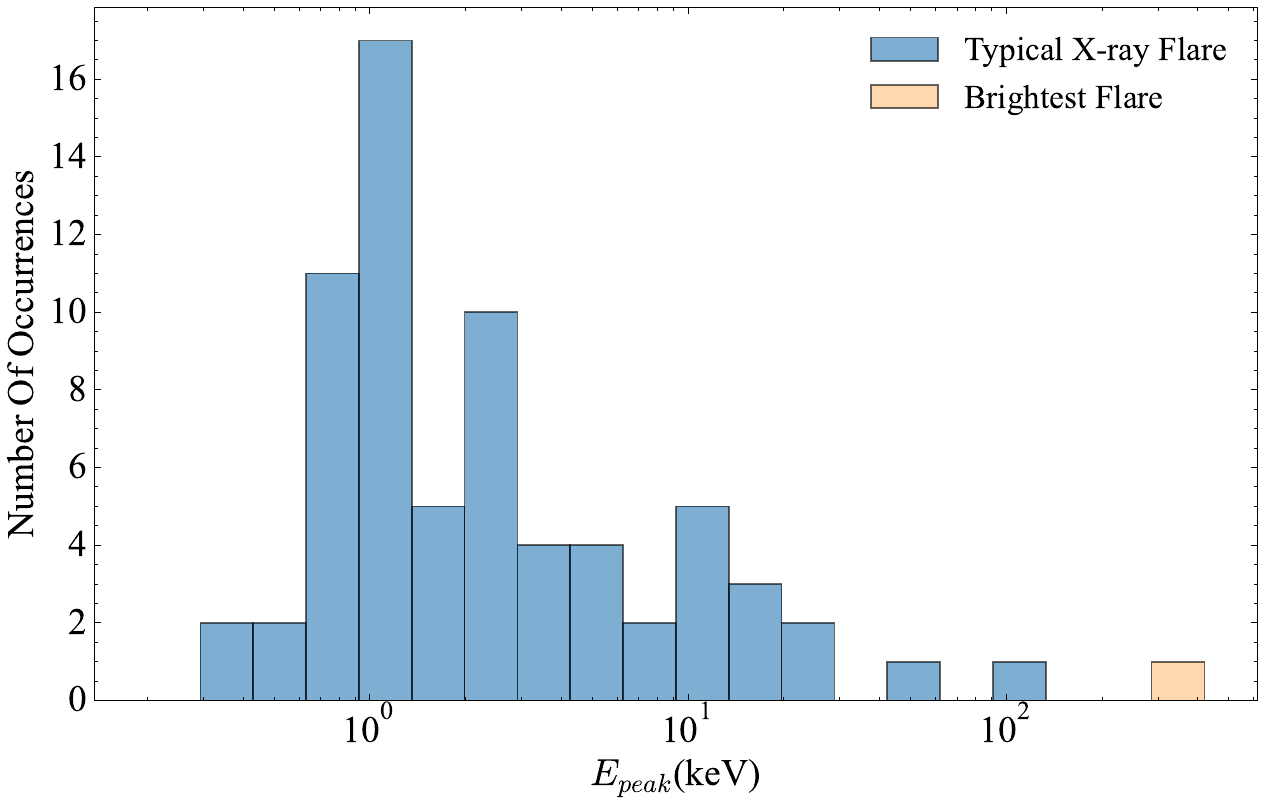}
        \subcaption{}
    \end{minipage}
\end{tabular}
  \caption{\label{fig_ratio_Ep}\small \textbf{Panel (a):} Blue dots represent typical X-ray flares fluence calculated in 0.2-10\,keV (data from \cite{A4_2007_flare_statis2}), red stars mark the Brightest Flare fluence calculated in 1-10000\,keV, the yellow line indicates the average fluence ratio (flare to prompt emission), and the red line shows the specific ratio for the Brightest Flare. \textbf{Panel (b):} Comparison of $E_{\text{peak}}$ between the  Brightest Flare and typical X-ray flares (data from \cite{A3_2007_flare_statis1,C7_2014_falre_sample,A14_2015_Mev_GeV}). The blue histogram represents the statistical distribution of $E_{\text{peak}}$ values for typical X-ray flares; the yellow histogram indicates the  Brightest Flare.}
\end{figure*}

The onset of the afterglow is generally defined as $T_{\rm AG} = T_0 + 225$\,s \citep{1_09A_Gecam,B9_2024_09A_afterglow}. By analyzing the afterglow data from joint observations of \textit{Fermi}/GBM and GECAM-C, we extended the afterglow light curve to cover the flare phase. The flare is clearly superimposed on a power-law decaying afterglow, as shown in Figure~\ref{fig_flux_all}. In this paper, smooth broken power-law (SBPL) functions (Eq.~\ref{eq_bpl}) \citep{A8_2016_Yi_flare_sample} is primarily applied to fit each flare of the GRB. For afterglow, the broken power-law function (BPL) (Eq.~\ref{eq_pl}) is primarily used \citep{B9_2024_09A_afterglow}. For the flare and afterglow, we also use a composite model to perform fit of the two phases (Eq.~\ref{eq_overall}).

\begin{equation}
F_{\text{flare,i}}(t) = F_{\rm 0i} \left[ \left( \frac{t}{t_{\rm ib}} \right)^{\alpha_{\rm i1} \omega} + \left( \frac{t}{t_{\rm ib}} \right)^{\alpha_{\rm i1} \omega} \right]^{-\frac{1}{\omega}}
\label{eq_bpl}
\end{equation}

where $\alpha_{\rm i1}$ and $\alpha_{\rm i2}$} are the temporal slopes of each flare, $t_{\rm ib}$ is the break time, and $\omega$ represents the sharpness of the peak of the component of the light curve. $F_{\text{flare,i}}(t)$ is the SBPL function fitted to each individual flare.

\begin{equation}
F_{\text{AG}}(t) = \begin{cases} 
F_{02}(t/t_b)^{-\alpha_3}, & t < t_b \\ 
F_{02}(t/t_b)^{-\alpha_4}, & t > t_b 
\end{cases}
\label{eq_pl}
\end{equation}

where $\alpha _3$ and $\alpha _4$ are the temporal slopes, $t_b$ is the break time, $F_{02}$ is the amplitude.

\begin{equation}
F(t) = \sum_{i=1}^{N} F_{\text{flare,i}}(t) + F_{\text{AG}}(t)
\label{eq_overall}
\end{equation}

$F(t)$ is a model that fits the flare phase and the afterglow phase simultaneously. The first term is the fit model for the entire flares obtained by summing the fit models for all individual flares. The second term is the model used to fit the afterglow.

The structure of the light curve indicates that the burst between $T_0 + 350$\,s and $T_0 + 600$\,s is a flare phase during the GRB afterglow phase. Furthermore, the multiple peaks in the light curve structure of the flare phase in Figure~\ref{fig_flux_all} also suggest that this BOAT flare is composed of multiple flares.

We conduct a detailed analysis of the brightest flare in the flare phase of GRB 221009A. We clarify that, in this work, the ``BOAT flare'' refers to the series of flares observed in GRB 221009A (from $T_0+350$\,s to $T_0+600$\,s), while the ``Brightest Flare'' specifically denotes the most intense flare within this series (from $T_0+500$\,s to $T_0+520$\,s). Unless otherwise specified, all subsequent calculations of parameters (such as fluence, $E_{\rm iso}$, $L_{\rm peak}$, etc.) pertain to this  Brightest Flare.

The fluence of the prompt emission is measured as $(1.20 \pm 0.01) \times 10^{-1}\,\rm{erg} \cdot \rm{cm}^{-2}$ in 1-10000\,keV \citep{1_09A_Gecam}, while the fluence of the  Brightest Flare is $(3.49 \pm 0.05) \times 10^{-3}\,\rm{erg} \cdot \rm{cm}^{-2}$ in 1-10000\,keV. The fluence ratio of the  Brightest Flare to the prompt emission is about 3\%. Previous statistical analyses have shown that the fluence of X-ray flares typically accounts for about 10\% of the fluence during the prompt emission phase \citep{A3_2007_flare_statis1, A4_2007_flare_statis2,C22_norris2}. The somewhat lower fluence ratio of the  Brightest Flare compared to this average value may suggest that the central engine may have already released the majority of its energy during the prompt emission (Figure~\ref{fig_ratio_Ep} Panel (a)).

The $T_{90}$ duration of GRB 221009A, calculated from either GBM or KW data, is all about 300\,s \citep{B82_kw_09A,A43_09A_GBM}. This value falls short of the typical $T_{90}$ definition of ultra-long GRBs, which is generally on the order of several thousand seconds \citep{A204_ULGRB}. Therefore, classifying GRB 221009A as an ultra-long GRB is disfavored \citep{A204_ULGRB,A203_ULGRB_09A,B82_kw_09A,A43_09A_GBM}.

For the emission episode occurring between $T_0$ + 350\,s and $T_0$ + 600\,s, we have previously presented multiple observational lines of evidence suggesting that it is more consistent with a GRB flare. These mainly include: the presence of a distinct afterglow component prior to the flare; a light-curve structure in which the flare is superimposed on an underlying power-law decaying afterglow and so on.  Collectively, these features support the interpretation of this emission episode as a GRB flare.

\begin{figure*}
  \centering
\begin{tabular}{cc}
  \begin{minipage}{0.42\textwidth}
      \includegraphics[width = \textwidth]{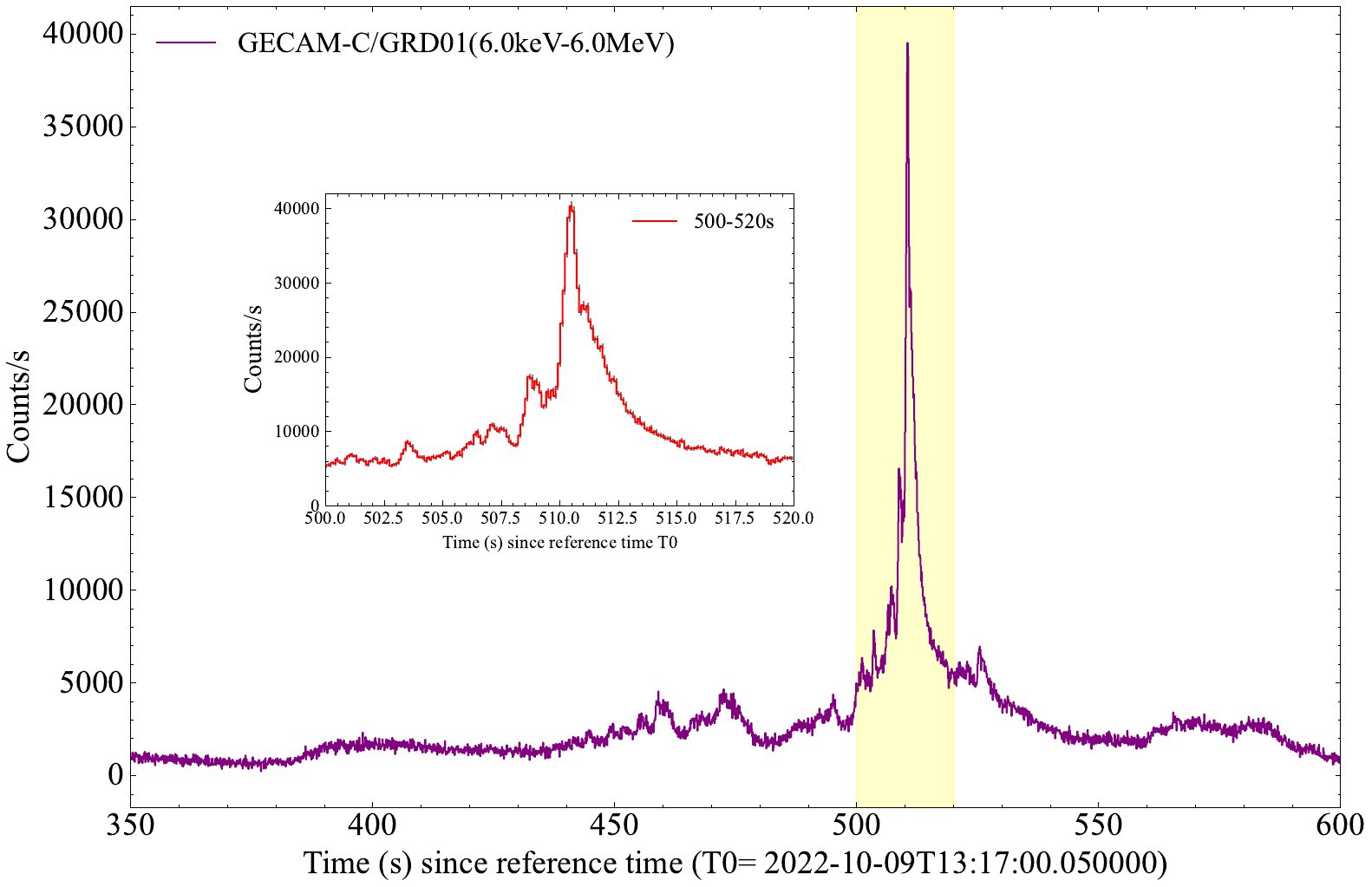}
      \subcaption{}
   \end{minipage}
    \begin{minipage}{0.45\textwidth}
        \includegraphics[width = \textwidth]{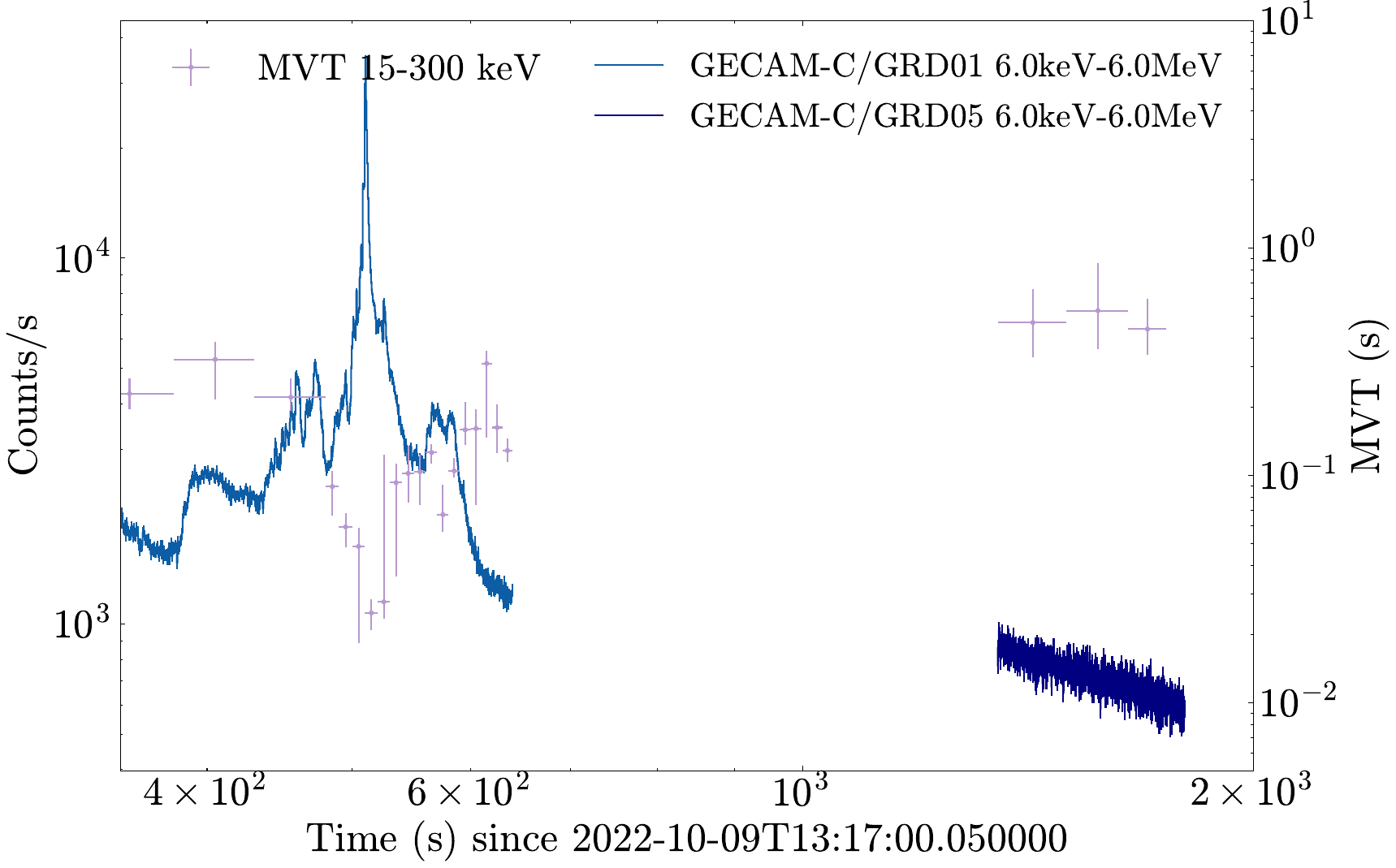}
        \subcaption{}
    \end{minipage}\\
    \begin{minipage}{0.9\textwidth}
        \includegraphics[width = \textwidth]{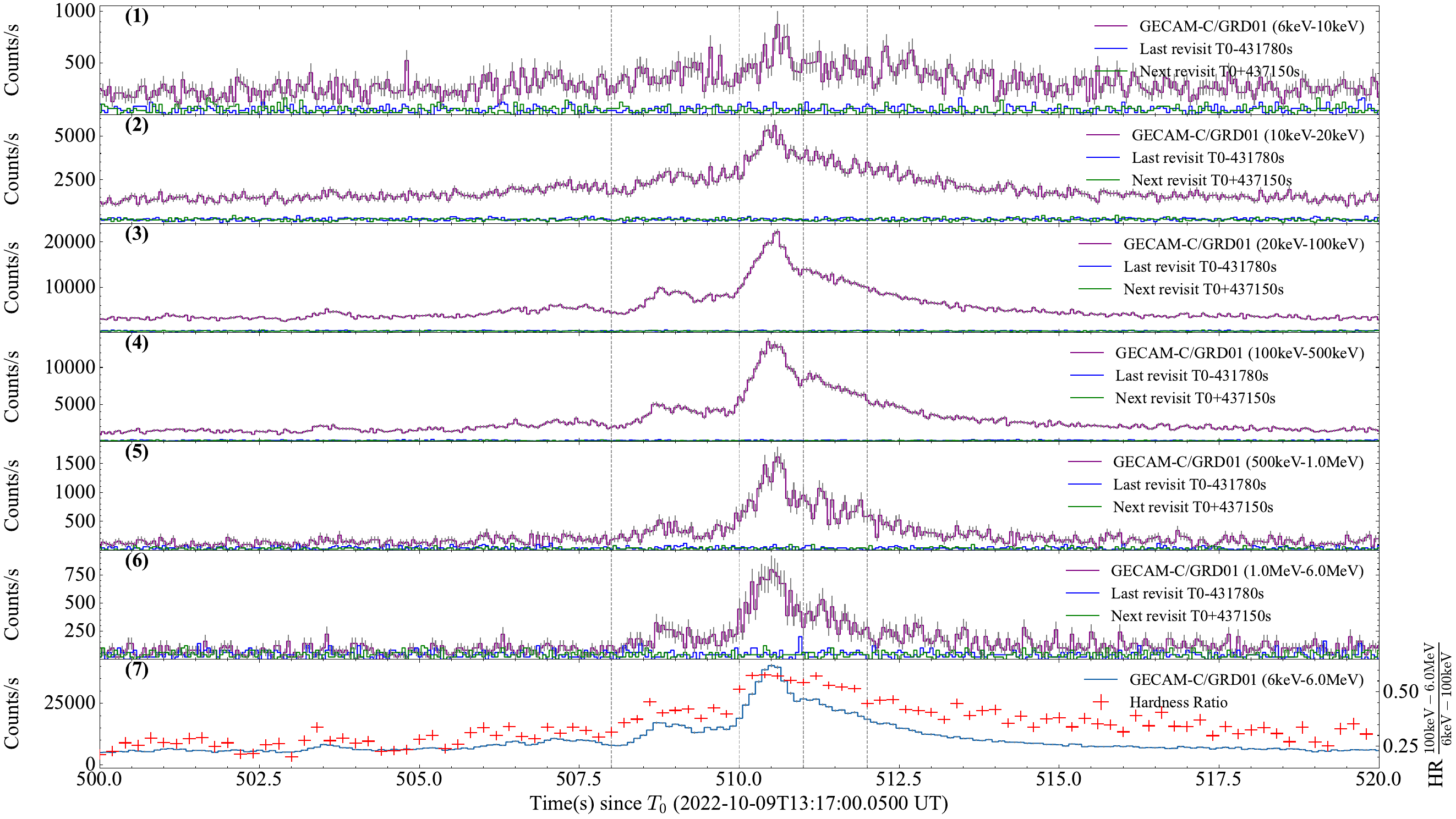}
        \subcaption{}
    \end{minipage}\\
\end{tabular}
  \caption{\label{fig_lc}\small \textbf{Panel (a):} Main: The purple line represents the net light curve from $T_0 + 350$\,s to $T_0 + 600$\,s, with the yellow-highlighted region indicating  Brightest Flare. \textbf{Inset:} The net light curve of the  Brightest Flare (from $T_0 + 500$\,s to $T_0 + 520$\,s). The time bin for net light curve is 0.1\,s. \textbf{Panel (b):} Evolution of the MVT is shown. The light and dark blue curves represent the light curves of the flare and afterglow phases observed by GECAM-C/GRD01 and GECAM-C/GRD05, respectively. The light purple data points indicate the calculated MVT values. \textbf{Panel (c):} Light curves for different energy bands and the evolution of spectral hardness of the  Brightest Flare as observed by GECAM-C/GRD01. \text{Panels (1) to (6):} The purple curves represent the light curves for the respective energy bands, while the blue and green lines indicate the background light curves derived from the revisited orbits. The time bin for all light curves is 0.05\,s. \text{Panel (7):} The blue line represents the light curve across the full energy band, with the red data points depicting the spectral hardness evolution.}
\end{figure*}

\subsection{Temporal analysis results}

\begin{figure*}
\centering
\begin{tabular}{cc}
\begin{minipage}{0.5\textwidth}
        \includegraphics[width = \textwidth]{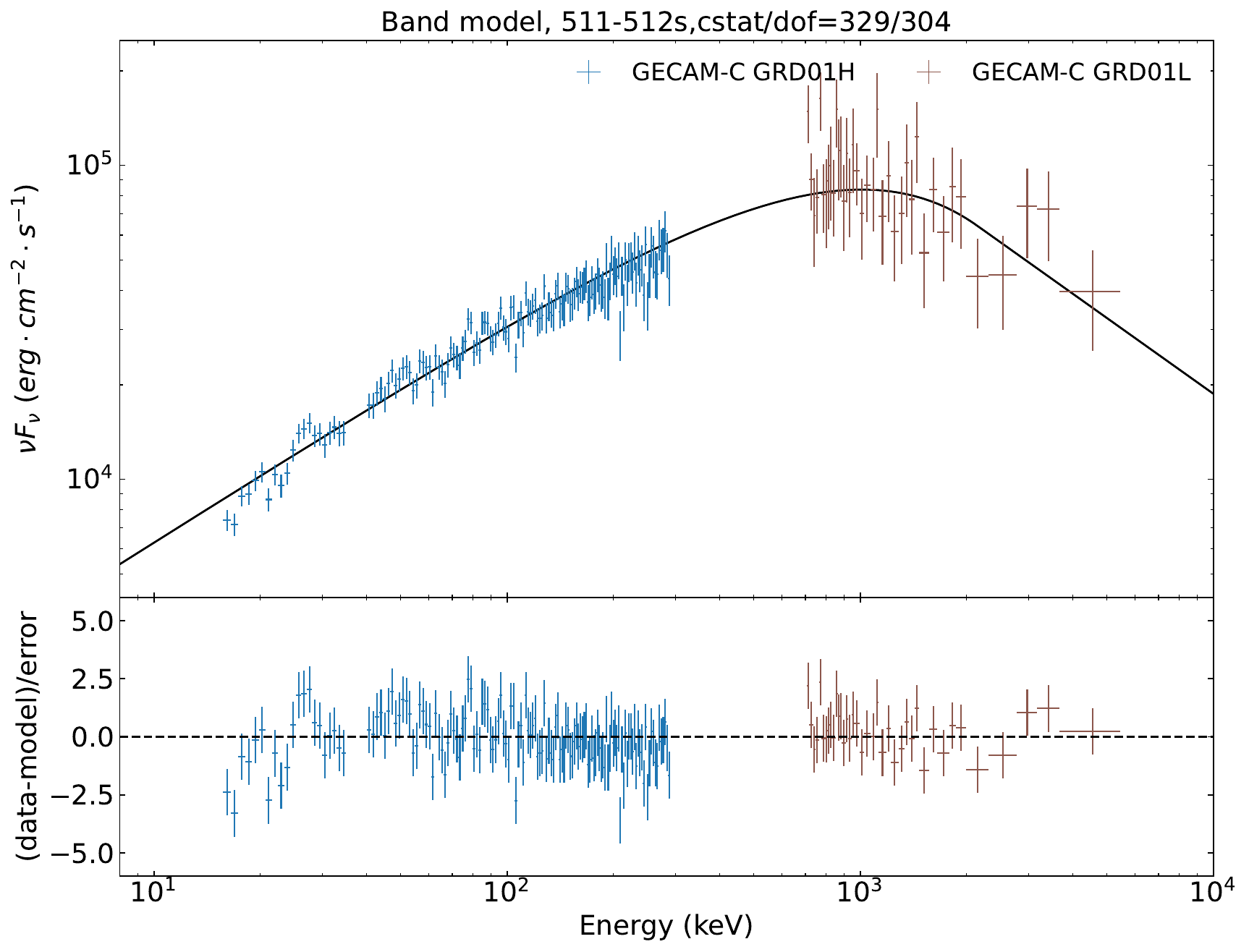}
        \subcaption{}
\end{minipage} 
\begin{minipage}{0.4\textwidth}
    \includegraphics[width = \textwidth]{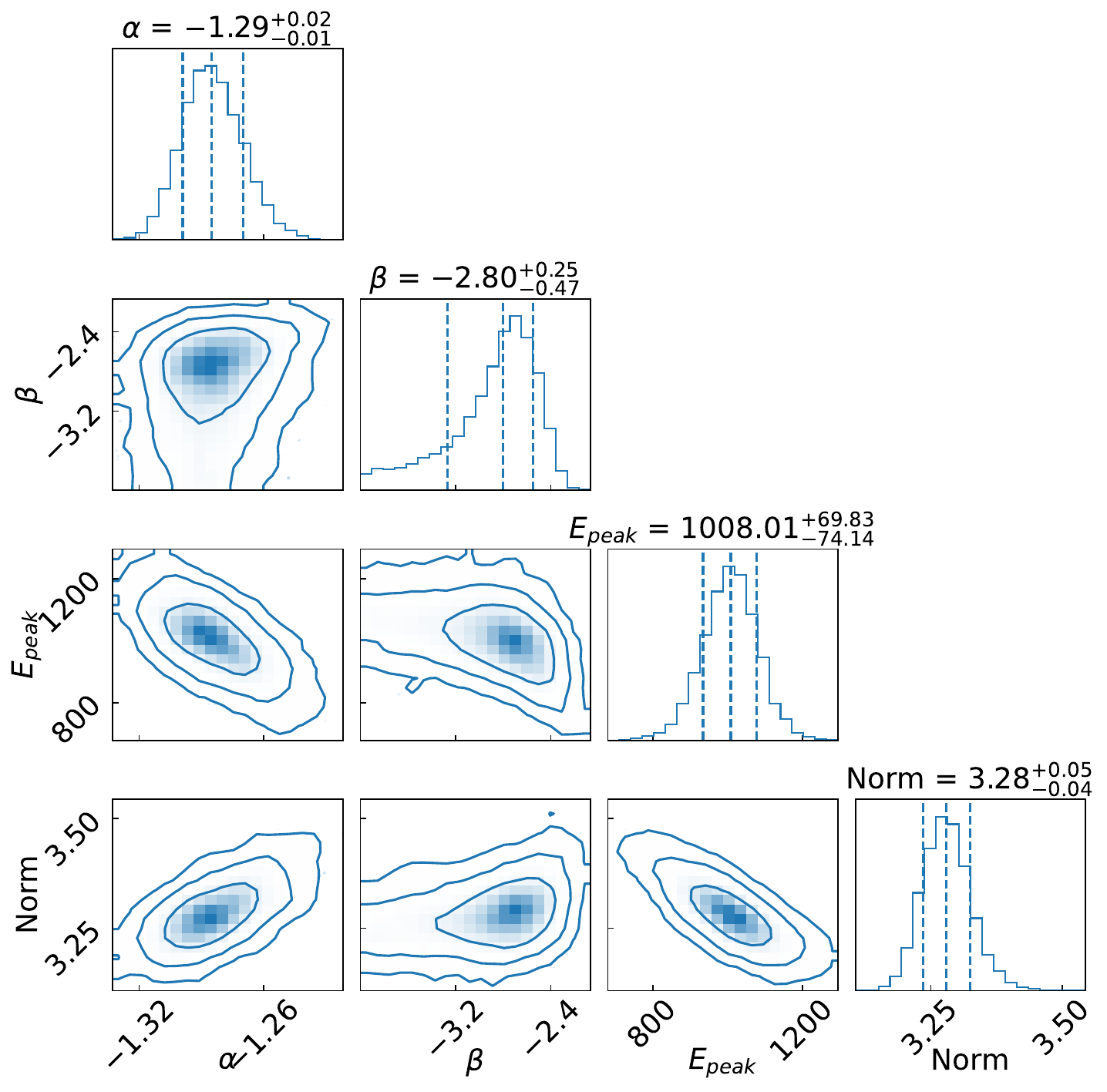}
    \subcaption{}
\end{minipage}\\
    \begin{minipage}{0.48\textwidth}
        \includegraphics[width = \textwidth]{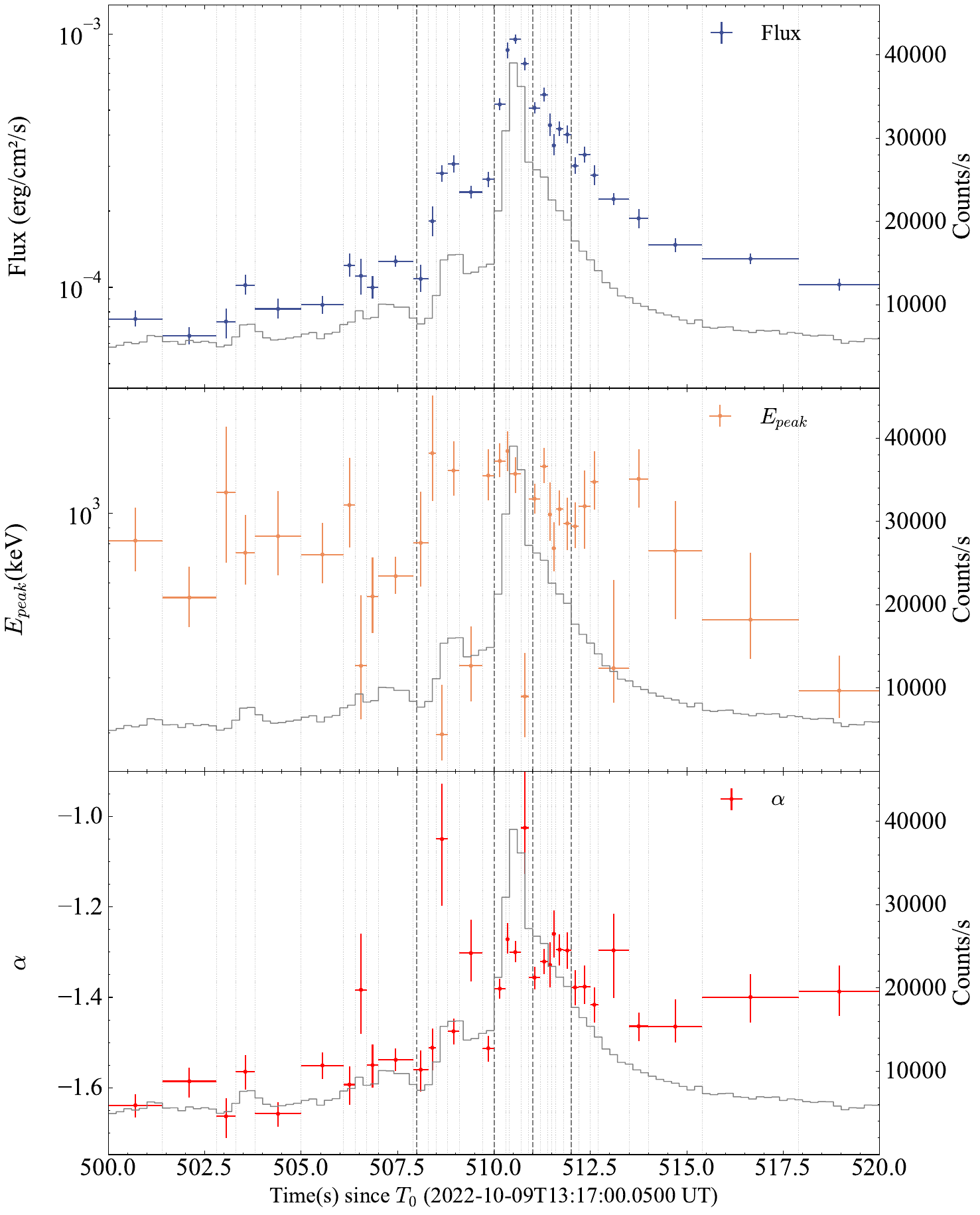}
        \subcaption{}
    \end{minipage}
    \begin{minipage}{0.47\textwidth}
        \includegraphics[width = \textwidth]{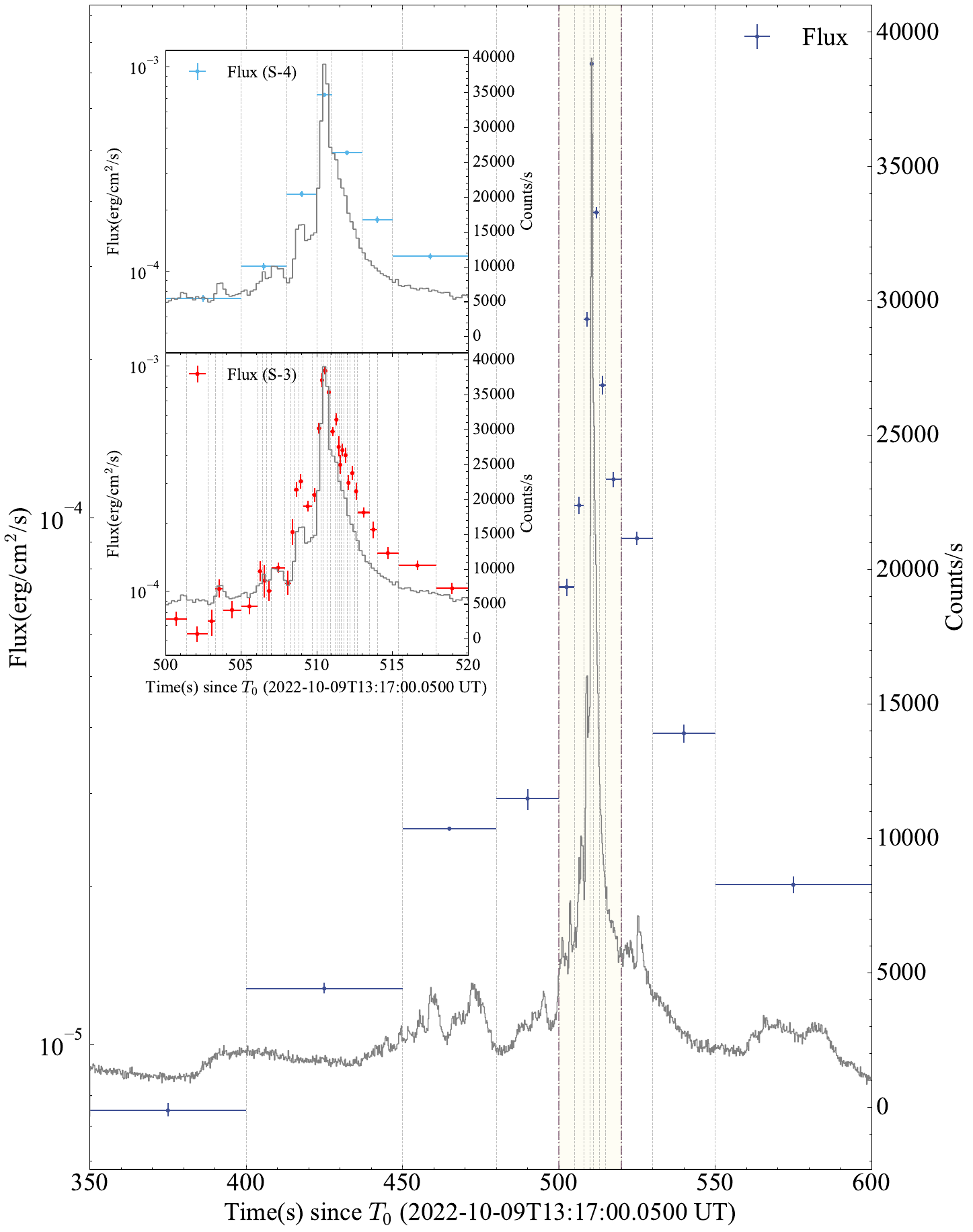}
        \subcaption{}
    \end{minipage}\\

\end{tabular}
  \caption{\label{fig4}\small  \textbf{Panel (a) and Panel (b):} The spectra of one time interval of the  Brightest Flare, along with their corresponding corner plots, are presented. The time periods analyzed, $T_0 + 511$\,s to $T_0 + 512$\,s. The spectra for this time interval includes the $\nu F_{\nu}$ spectra and the residual maps. The Band function was employed to fit this time interval. \textbf{Panel (c):} Evolution of the spectral parameters from the spectrum fit to S-3 over time. The three subplots, arranged from top to bottom, correspond to the evolution of flux, $E_{\text{peak}}$, and $\alpha$. The gray curve represents the light curve of the BOAT flare. \textbf{Panel (d):} Main flux transition for the wider time bins of S-4, spanning $T_0 + 350$\,s to $T_0 + 600$\,s, with the  Brightest Flare highlighted by a light yellow background. The two subplots in the smaller panel on the left display the flux transitions during the main phases of S-4 and S-3, arranged from top to bottom.}
\end{figure*}

The light curve of the GRB 221009A flare is shown in Figure~\ref{fig_lc} and Figure~\ref{fig4}. The BOAT flare exhibits an extremely high flux, exceeding that of most GRBs \citep{1_09A_Gecam}, and an apparently different structure in its light curve compared to other typical flares observed in X-ray or optical band \citep{A1_2006_flare_physics,A3_2007_flare_statis1,A4_2007_flare_statis2,A25_optical_flare,C7_2014_falre_sample,A14_2015_Mev_GeV}. 

Most flares, particularly X-ray flares detected by the \textit{Swift}/XRT, typically exhibit a simple structure characterized by a rapid rise followed by a decline \citep{A1_2006_flare_physics,A3_2007_flare_statis1,A4_2007_flare_statis2,A8_2016_Yi_flare_sample}. In contrast, the light curve of the BOAT flare displays a complex pattern with many small pulses. Detection of such fine features of the flare should be attributed to the exceptional brightness of this flare. This intricate structure is a superposition of multiple pulses, resembling the light curves observed during the prompt emission of GRBs. Due to the relatively low brightness of typical flares, the high temporal resolution (0.1\,s time-bin) multipulse superimposed light curve structure observed here has rarely been detected before. Rapid variability timescales of the flare light curve strongly exclude an external shock origin. Fast variability of the flare suggests that both the flare and the prompt emission have the same physical mechanism: they all likely originate from the activities of the central engine.

The brightest phase of the  Brightest Flare is mainly concentrated between $T_0 + 508$\,s and $T_0 + 512$\,s, with notable peaks observed at $T_0 + 509$\,s, $T_0 + 511$\,s, and $T_0 + 512$\,s. The brightest peak occurs near $T_0 + 511$\,s (Figure~\ref{fig_lc}). In the 500\,keV-1\,MeV energy range, the flare exhibits a less prominent double-peak structure at the peak time, while other energy ranges display a single pulse with a broader temporal structure.

The light curves of  Brightest Flare across different energy bands reveal a clear and complete structure spanning from the X-ray band to the soft $\gamma$-ray band. The majority of photons from the  Brightest Flare are concentrated in the hard X-ray band of 20\,keV--100\,keV. Notably, in the MeV band, the  Brightest Flare exhibits a distinct and complete structure. The  Brightest Flare has a photon count rate comparable to that of the main emission of most GRBs. This marks the first observation of the high statistics and high time resolution of a flare light curve in the keV-MeV band \citep{A14_2015_Mev_GeV}. The  Brightest Flare provides an opportunity to reveal high-precision spectral evolution, making the evolution of the spectral hardness more concrete during the flare.

Photons in the MeV band are predominantly concentrated between $T_0 + 508$\,s and $T_0 + 515$\,s, with a peak count rate exceeding 700 counts/s. 
This observation allows us to conduct a detailed analysis of the flare process during the afterglow phase in the $\gamma$-ray band. The light curve in the $\gamma$-ray band suggests that this flare may represent a rarely detected and confirmed $\gamma$-ray flare. This  Brightest Flare exhibits a rapidly varying light curve structure, similar to that of the prompt emission in the same energy range (6\,keV-6\,MeV). This similarity provides direct observational evidence that the flare and prompt emission share a common origin.

Panel (7) of Figure~\ref{fig_lc} Panel (c) shows the temporal evolution of the spectral hardness of the  Brightest Flare. With fine time resolution, we reveal spectral hardening during multiple pulses in the light curve. Notably, we find that the spectral hardness of the  Brightest Flare is positively correlated with the flux.

In the prompt emission of GRBs, there are two recognized trends in the evolution of spectral hardness. The first trend indicates that the spectral hardness gradually decreases over time \citep{C1_1996_grb_spec_evolution1, C2_2010_grd_spec_evolutio2}. The second trend is that the spectral hardness is positively correlated to the photon flux \citep{C3_2007_flux1, C4_2010_flux2}. Through precise temporal and spectral measurements, we find that this relationship persists even during most of the pulses in the flare. This behavior is closely related to the evolution of the spectral hardness typically observed in prompt emission \citep{A1_2006_flare_physics,A3_2007_flare_statis1,A4_2007_flare_statis2}. Our findings provide strong evidence that the  Brightest Flare follows the same spectral evolution pattern as the prompt emission of GRBs.

We also present the Minimum Variability Timescale (MVT) calculations in 15-300\,keV for both the BOAT flare and subsequent afterglow phases (Figure~\ref{fig_lc}). For the afterglow phase, the data from GRD05 were used to compute the MVT, owing to its smaller incident angle and superior data quality compared to other detectors. The results indicate that the MVT decreases significantly during the flare, reaching a minimum value of about 0.018\,s. The MVT during the afterglow phase remains around 0.5\,s. Our analysis demonstrates that the MVT derived from GECAM-C is consistent with that provided by \textit{Fermi}/GBM \citep{A43_09A_GBM,A42_09A_insight}. Furthermore, GECAM-C successfully reconstructs the MVT during time intervals where \textit{Fermi}/GBM data are affected by instrumental effects.

\subsection{Spectral analysis results}

For the analysis of the spectrum of all time periods of the BOAT flare, the selected energy bands were 15-300\,keV for the HG and 0.7-5.5\,MeV for the LG. The selection of this energy band has been demonstrated to be well suited for the analysis of GRB 221009A in previous studies \citep{A104_C_cross,1_09A_Gecam,B10_2024_09A_line}.

To characterize the overall spectral properties of the BOAT flare, we initially used the wide time interval S-1. During the  Brightest Flare (from $T_0 + 500$\,s to $T_0 + 520$\,s), and intervals outside the Brightest Flare (from $T_0 + 350$\,s to $T_0 + 500$\,s and from $T_0 + 520$\,s to $T_0 + 600$\,s), we all use a Band function. The results of spectral adjustment for the wide time frame S-1 are detailed in Table~\ref{table_spec}.

To further analyze the spectral properties of the  Brightest Flare, this phase was divided into finer time intervals (i.e. S-2). Significant peaks were observed near $T_0 + 509$\,s, $T_0 + 510$\,s, and $T_0 + 512$\,s. Therefore, a spectral adjustment was performed for these time intervals. We chose to fit the spectra using the Band function or CPL model for all the bins of S-2. The spectral fitting result for one selected time interval and its corner plots are shown in Figure~\ref{fig4} Panel (a). The residuals of the spectral fits, shown in Figure~\ref{fig4} Panel (a), do not exhibit significant evolutionary structures, indicating good fitting performance. The spectral fitting results for the  Brightest Flare under finer time intervals (S-2) are summarized in Table~\ref{table_spec}.

GECAM-C provided precise measurements of the BOAT flare with high temporal resolution, allowing us to perform high resolution time binning for the  Brightest Flare to do spectral analysis (i.e. S-3). We performed refined temporal binning using the Bayesian Block \citep{C20_BB}. Under the S-3 time binning scheme, we have for the first time performed time-resolved spectral fitting of GRB flares within narrow time slices of 0.1\,s. The results of the high precision time resolved S-3 analysis are presented in Table~\ref{table_spec}. Given the relatively random evolution of the blackbody component parameters during the flare phase, we suggest that a significant thermal component is unlikely to be present throughout the flare.

Table~\ref{table_spec} presents the spectral fitting results for each time interval of the BOAT flare. The parameter $E_{\text{peak}}$ denotes the peak energy of the $\nu F_{\nu}$ spectrum. Using the low-energy spectral index $\alpha$ and the characteristic energy $E_{\text{cut}}$ obtained from the spectral fitting, the peak energy can be derived using the formula $E_{\text{peak}} = (\alpha + 2) E_{\text{cut}}$. Fluence is calculated in 1-10000\,keV. The ratio of the c-statistic to the degrees of freedom (\texttt{cstat}/dof) is provided to assess the goodness of fit.

Figure.~\ref{fig4} (c) shows the evolution of the spectral parameters fitted to the  Brightest Flare. The evolution becomes more pronounced as the flux of the  Brightest Flare rapidly increases. Regarding spectral parameters, the low-energy indices $\alpha$ and $E_{\text{peak}}$ of the  Brightest Flare exhibit a hardening trend as the flux increases. In contrast, as the  Brightest Flare diminishes, $\alpha$ and $E_{\text{peak}}$ tend to soften. This spectral evolution indicates that the flare generation mechanism is similar to that of prompt emission, which is consistent with previous studies \citep{A1_2006_flare_physics,A3_2007_flare_statis1,A4_2007_flare_statis2}.

Figure~\ref{fig4} (d) shows a comparison of the flux light curves with different time resolutions (S-3, S-4). In the time interval from $T_0 + 350$\,s to $T_0 + 600$\,s, the lower time resolution S-4 (see the S-4 time slice in Table~\ref{table_spec}) displays a rapid rise followed by a decline, resembling the structure of a typical flare. However, when we perform spectral fitting using the higher time resolution slice S-3, the flux light curve still retains the pattern of rapid rise and decline, and it also shows three peaks that are consistent with those in the resolved light curves of different energy bands in Figure~\ref{fig_lc}.

The second subplot in Figure~\ref{fig4} (c) shows the temporal evolution of $E_{\text{peak}}$ during the Brightest Flare. The general evolution of $E_{\text{peak}}$ shows an initial upward trend followed by a downward trend. A comprehensive spectral fit to the  Brightest Flare shows that $E_{\text{peak}}$ is about 300 keV. The third subplot in Figure~\ref{fig4} (a) shows the temporal evolution of the spectral index $\alpha$. Each peak in the light curve corresponds to a noticeable variation in $\alpha$, which reveals curves similar to those observed in the prompt emission. We note that the maximum value of $\alpha$ reaches approximately $-1$, and there is no period in which $\alpha$ exceeds the synchrotron death line of $-2/3$ \citep{C6_1994_alpha_2/3,C8_1998_2/3_2}.

During the  Brightest Flare, the fitted values of $E_{\text{peak}}$ consistently remain within the $\gamma$-ray energy band. Notably, this observation represents the first detection of $E_{\text{peak}}$ within the $\gamma$-ray energy band during flares phase of a GRB \citep{A3_2007_flare_statis1,A14_2015_Mev_GeV}. The light curve in the MeV band and the $E_{\text{peak}}$ in the $\gamma$-ray band of the  Brightest Flare provide evidence that it is a rare $\gamma$-ray flare.

 Previous studies have shown that spectra of typical X-ray flares samples observed by \textit{Swift} can be fitted with the Band function and the CPL model \citep{A3_2007_flare_statis1,A14_2015_Mev_GeV,C7_2014_falre_sample}. These samples can provide the parameter $E_{\text{peak}}$ for statistical analysis. In our analysis, we compared the $E_{\text{peak}}$ of the  Brightest Flare with that of typical X-ray flares. Our results indicate that the $E_{\text{peak}}$ of typical X-ray flares is mainly concentrated in the soft X-ray band, usually around 1\,keV. In samples collected from previous studies, the highest $E_{\text{peak}}$ of typical X-ray flares was 99\,keV \citep{A3_2007_flare_statis1,A14_2015_Mev_GeV,C7_2014_falre_sample}. Therefore, this flare has the highest peak energy $E_{\text{peak}}$ detected so far (Figure~\ref{fig_ratio_Ep} Panel (a)).

\subsection{Comparison study with typical X-ray flares}

Table~\ref{table2} presents the parameters of the  Brightest Flare, as well as the corresponding parameter ranges for typical X-ray flares. For comparison, we used a sample of X-ray flares from a previous study \citep{A8_2016_Yi_flare_sample}. Specifically, the  Brightest Flare exhibits the highest isotropic energy $E_{\text{iso}}$, peak luminosity $L_{\text{peak}}$, isotropic luminosity $L_{\text{iso}}$, and fluence among all known flares.

Due to the complex structures around the  Brightest Flare, it is challenging to fit the timescale for the entire  Brightest Flare. Given the prior practice of fitting only single pulses for rise/decay timescales, we exclusively analyze the brightest pulse for the timescale. The duration $\Delta t$ of the flare can be determined by calculating the $T_{90}$ (the time interval during which 5\% to 95\% of the total number of photons are accumulated) during the  Brightest Flare from $T_0$+500\,s to $T_0$+520\,s. In this paper, we use Bayesian Block to calculate $T_{90}$ \citep{C20_BB}. The Norris05 function is used to fit the peak portion of the flares \citep{C21_norris1,C22_norris2}. The formula for the Norris05 function is shown in Eq.~\ref{eq3}. We calculated the $T_{90}$, Fluence, $E_{\text{iso}}$ and $L_{\text{iso}}$ of the brightest pulse within Norris05 function fitted time interval of the Brightest Flare \citep{C21_norris1,C22_norris2}. The fitting results of the Norris05 function are shown in Figure~\ref{figA1}.

\begin{equation}
\mathrm{C}(t) = A \lambda \exp \left( -\frac{\tau_{1}}{(t - t_{\mathrm{s}})} - \frac{(t - t_{\mathrm{s}})}{\tau_{2}} \right) \quad \text{for } t > t_{\mathrm{s}}
\label{eq3}
\end{equation}

where $\mu = (\tau_1 / \tau_2)^{1/2}$ and $\lambda = e ^ {2 \mu}$. The peak time of the Norris05 function is shown in Eq.~\ref{eq4}. 

\begin{equation}
T_{\mathrm{pk}} = \tau_{\mathrm{pk}} + t_{\mathrm{s}} = \left( \tau_{1} \tau_{2} \right)^{1/2} + t_{\mathrm{s}}
\label{eq4}
\end{equation}

The pulse width is measured between the two points, $\omega = \Delta t_{1/e} = t_{\rm decay} +t_{\rm rise}= \tau_2(1+4 \mu)^{1/2}$. The pulse asymmetry is $k = \frac{t_{\rm decay} - t_{\rm rise}}{t_{\rm decay} + t_{\rm rise}} = ( 1 + 4\mu )^{-1/2}$. The rise time $T_{\rm rise}$ and the decay time $T_{\rm decay}$ are expressed in terms of $\omega$ and $k$ as in Eq.~\ref{eq5}.

\begin{equation}
t_{\mathrm{decay, rise}} = \frac{1}{2} \omega(1 
 \pm k)
\label{eq5}
\end{equation}

By fitting the Norris05 function, we derive $T_{\rm rise}$, $T_{\rm decay}$, and the peak time $T_{\rm peak}$. The calculation formula is shown in Eq.~\ref{eq4} and Eq.~\ref{eq5}. It is important to note that the result of $T_{\rm pk}$ is calculated from $T_0$. The precursor of GRB 221009A, which is triggered at $T_0$, has a long waiting time before the main emission, which is very rare in GRBs. Most GRBs are triggered starting from the main emissions. For a fair comparison, the $T_{\rm peak}$ of the  Brightest Flare should be calculated from the beginning of the main emission, starting from $T_0 + 215$\,s. Therefore, $T_{\rm peak}$ = $T_{\rm pk}$ - 215\,s. Notably, since this paper only calculates the $T_{\rm rise}$ and $T_{\rm decay}$ for the brightest pulse of the  Brightest Flare, the sum of $T_{\rm rise}$ and $T_{\rm decay}$ significantly differs from the duration. However, this does not affect the final conclusions.

\begin{figure*}[htbp]
  \centering
\begin{tabular}{cc}
  \begin{minipage}{0.4\textwidth}
      \includegraphics[width = \textwidth]{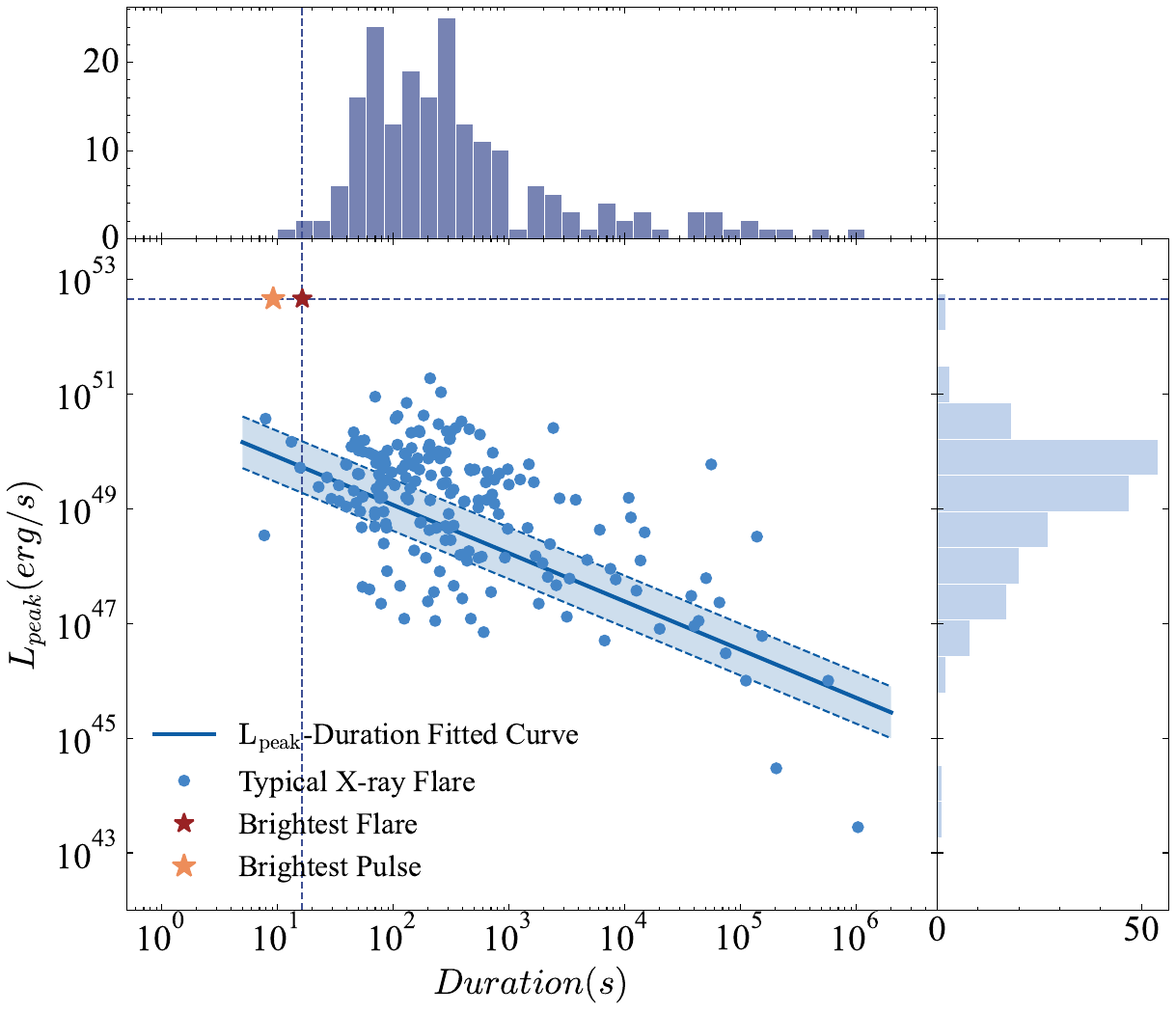}
      \subcaption{}
   \end{minipage}
    \begin{minipage}{0.4\textwidth}
        \includegraphics[width = \textwidth]{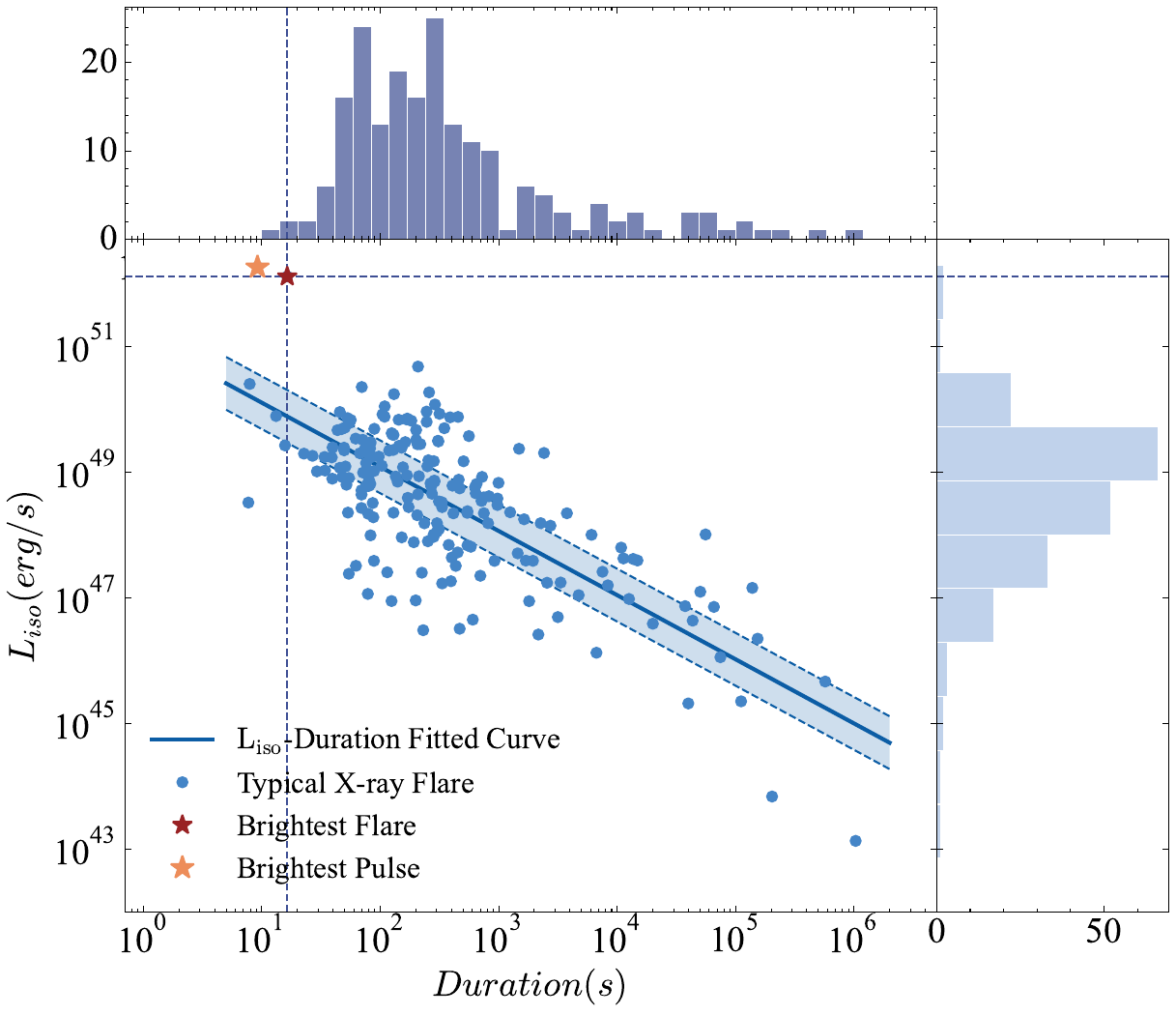}
        \subcaption{}
    \end{minipage}\\
    \begin{minipage}{0.4\textwidth}
        \includegraphics[width = \textwidth]{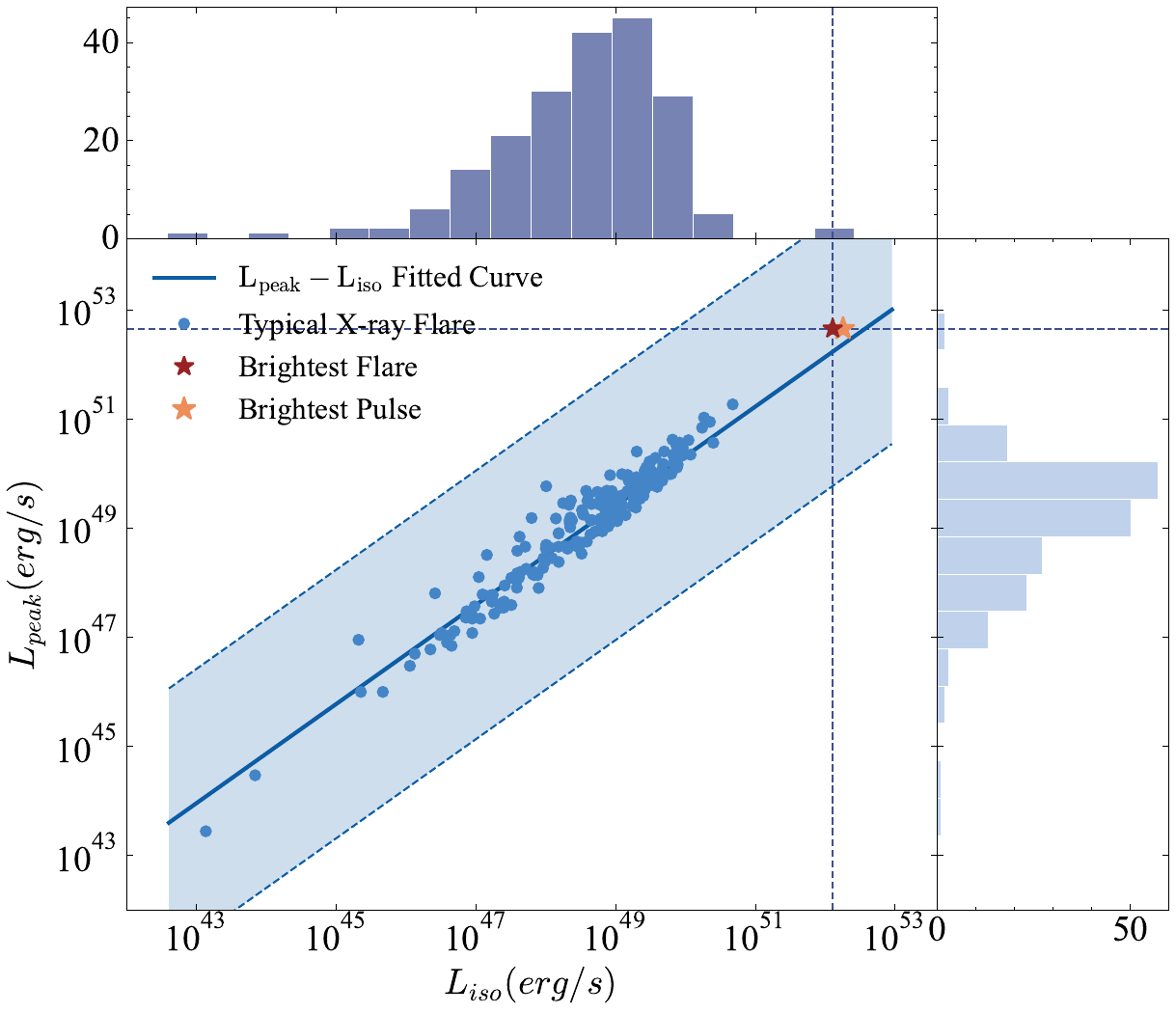}
        \subcaption{}
    \end{minipage}
    \begin{minipage}{0.4\textwidth}
        \includegraphics[width = \textwidth]{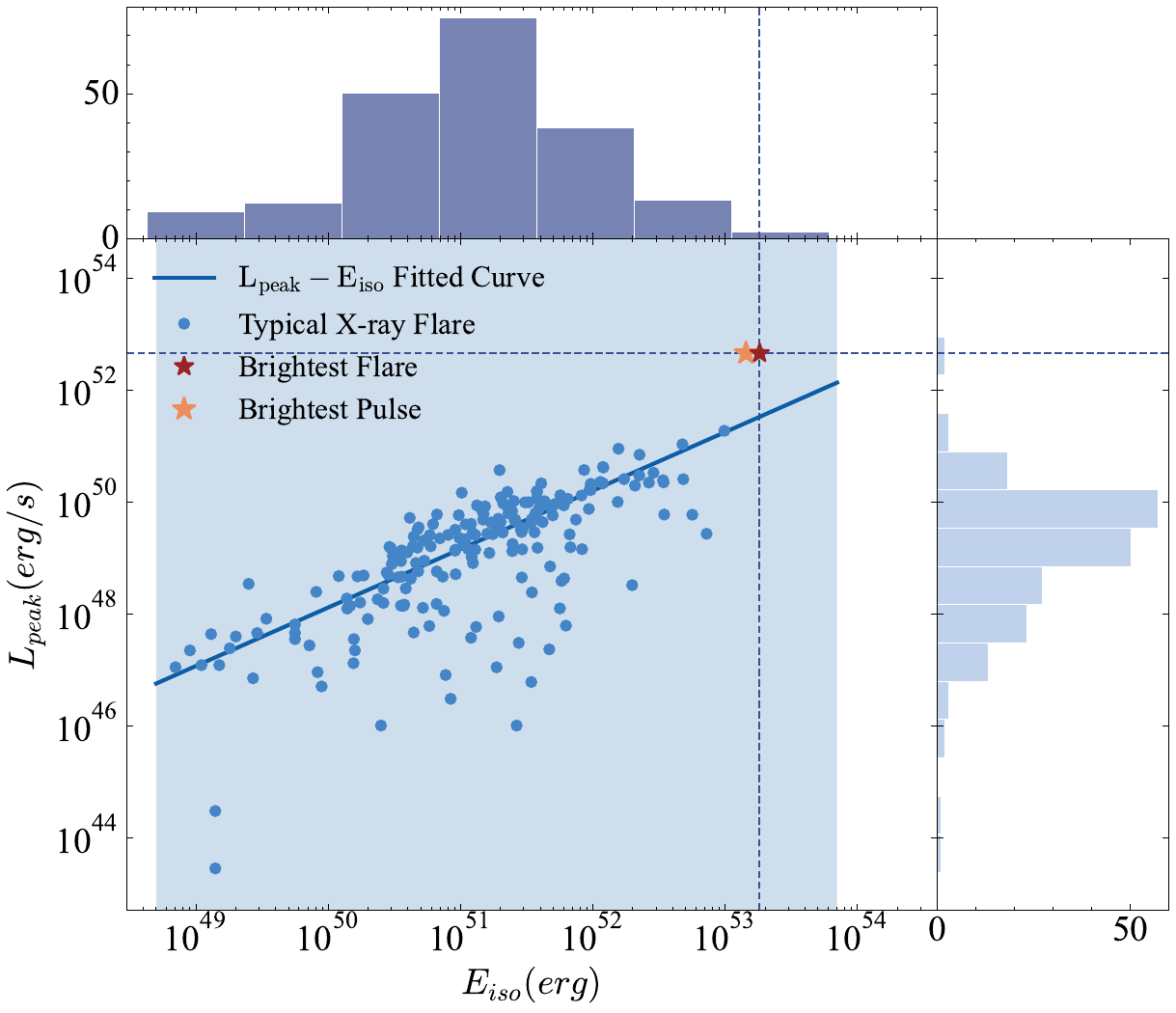}
        \subcaption{}
    \end{minipage}\\
    \begin{minipage}{0.4\textwidth}
        \includegraphics[width = \textwidth]{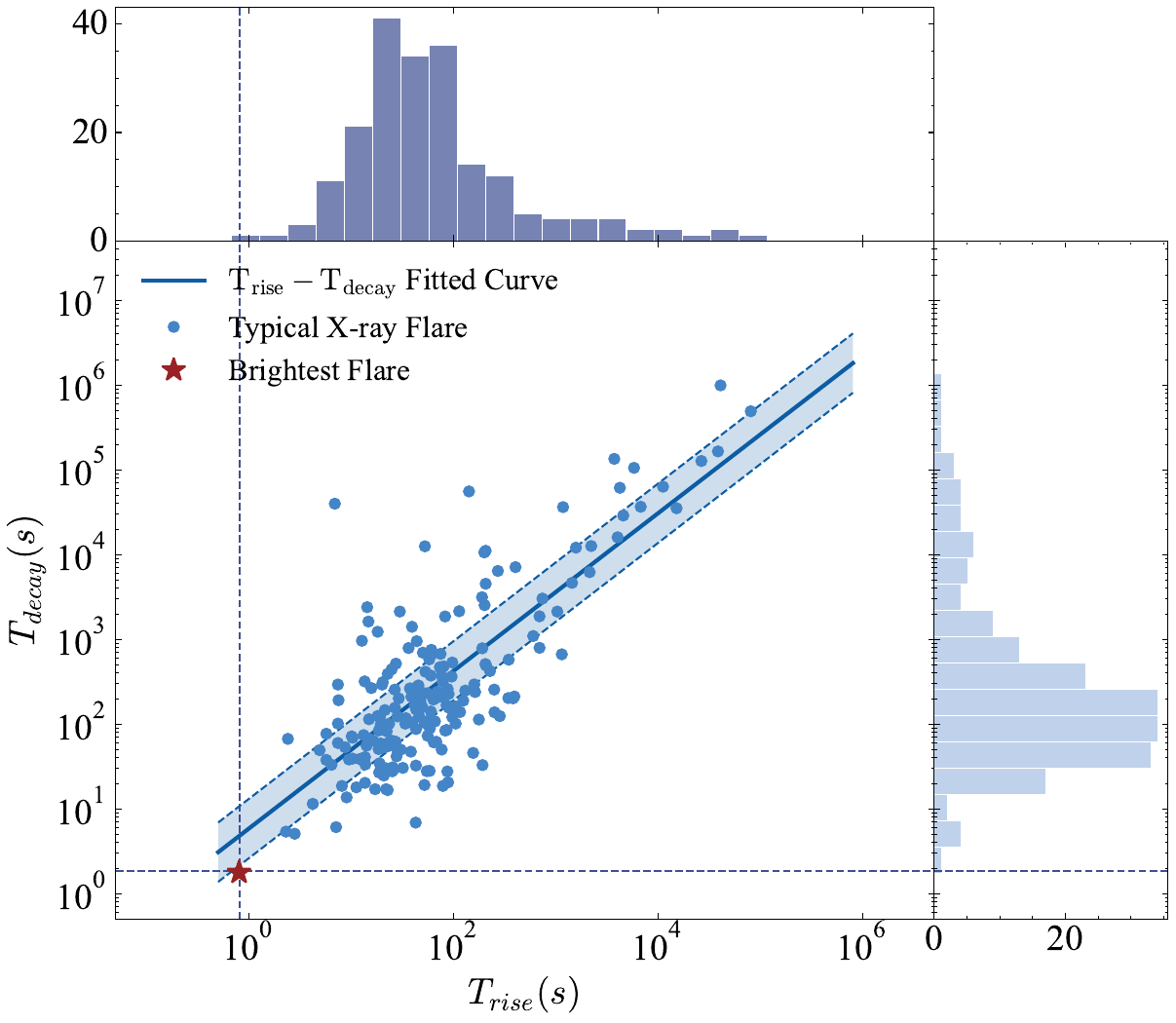}
        \subcaption{}
    \end{minipage}
    \begin{minipage}{0.4\textwidth}
        \includegraphics[width = \textwidth]{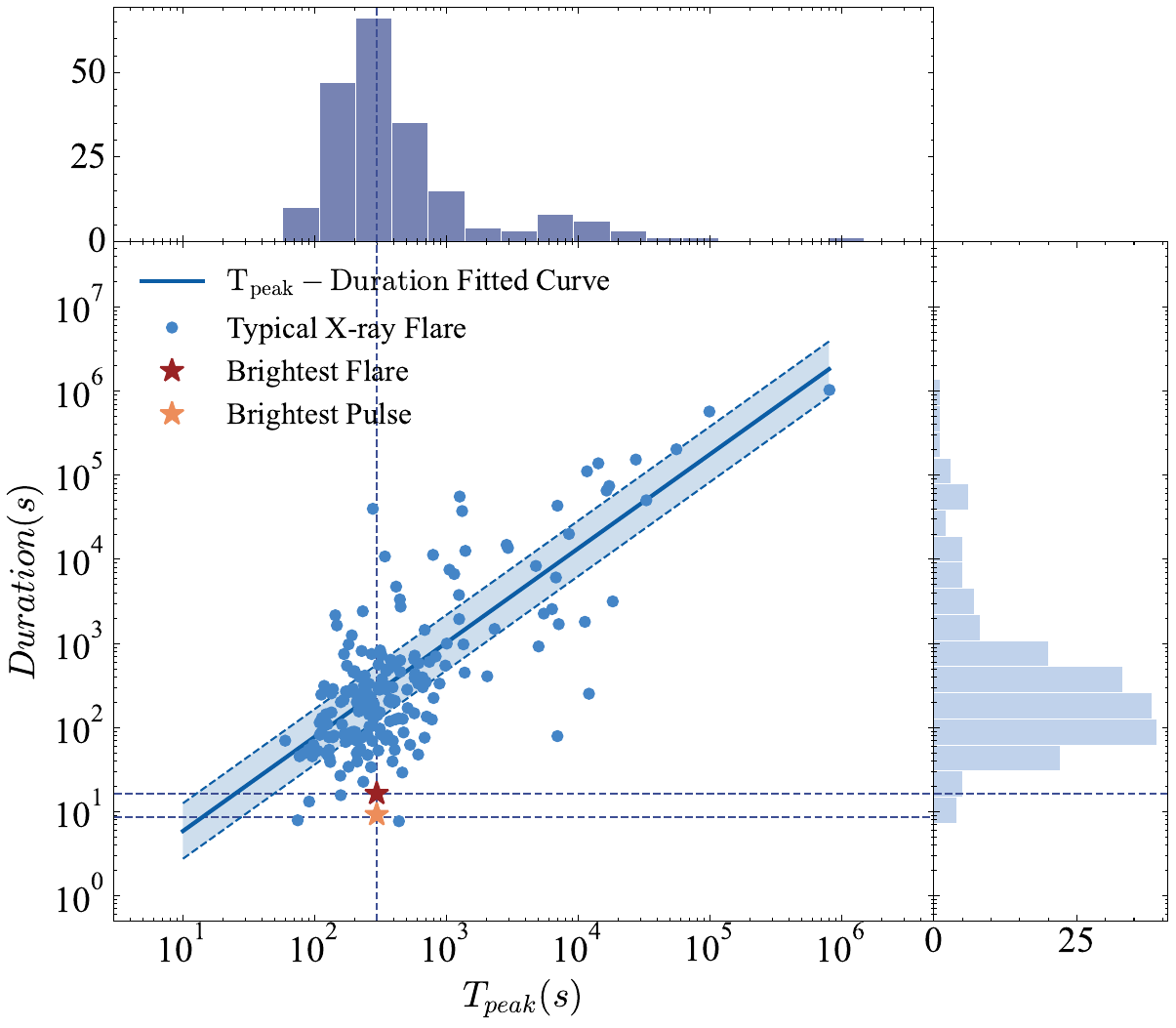}
        \subcaption{}
    \end{minipage}\\
\end{tabular}
  \caption{\label{fig6}\small This figure presents the statistical distributions of the parameters for the  Brightest Flare in comparison to those of typical X-ray flares. Panels (a) to (f) show the distributions for $L_{\text{peak}}$--Duration, $L_{\text{iso}}$--Duration, $L_{\text{peak}}$--$L_{\text{iso}}$, $L_{\text{peak}}$--$E_{\text{iso}}$, $T_{\text{decay}}$--$T_{\text{rise}}$ and Duration--$T_{\text{peak}}$ respectively. The histograms on the upper and right sides of each subplot represent the statistical distributions of the parameters corresponding to the horizontal and vertical axes, respectively. The blue data points denote typical X-ray flares, while the red pentagrams mark the  Brightest Flare and the orange pentagrams mark the brightest pulse of the Brightest Flare. The dark blue straight line indicates the results of the statistical analysis \citep{A8_2016_Yi_flare_sample}. The blue region is the 3 $\sigma$ uncertainty regions for the best fit line.}
\end{figure*}

The $E_{\text{iso}}$ can be obtained by $E_{\text{iso}} = 4\pi D_{\mathrm{L}}^2 S_{\mathrm{F}} / (1+z)$, where $z$ is the redshift, $D_{\mathrm{L}}$ is the luminosity distance, and $S_{\mathrm{F}}$ is the fluence, which is obtained by time integrating the spectrum of the  Brightest Flare in the 1-10000\,keV. The $L_{\text{iso}}$ is derivated from $L_{\text{iso}} = (1+z)E_{\text{iso}}/\Delta t$, while the peak luminosity follows $L_{\text{peak}} = 4\pi D_{\mathrm{L}}^2 F_{\text{peak}} / (1+z)$, where $F_{\text{peak}}$ is the peak flux. The peak flux is obtained by calculating the flux during the peak time interval from $T_0$+510\,s to $T_0$+511\,s.

As shown in Table~\ref{table2}, the  Brightest Flare is significantly brighter than typical X-ray flares, with $L_{\text{peak}}$, $L_{\text{iso}}$, and $E_{\text{iso}}$ exceeding those of typical X-ray flares by 2 to 3 orders of magnitude. To explore the statistical relationships among these parameters, we analyzed 6 pairs of key parameters: $L_{\text{peak}}$--Duration, $L_{\text{peak}}$--$E_{\text{iso}}$, $L_{\text{peak}}$--$L_{\text{iso}}$, $L_{\text{iso}}$--Duration, $T_{\text{decay}}$--$T_{\text{rise}}$ and Duration--$T_{\text{peak}}$. The results of this comparative analysis are presented in Figure~\ref{fig6}. Since the $T_{\text{rise}}$, $T_{\text{decay}}$, $T_{\text{peak}}$, and duration of the  Brightest Flare are not significantly different from those of typical X-ray flares, the statistical relationships between the timescales ($T_{\text{rise}}$ and $T_{\text{decay}}$) and energy mentioned above are not provided separately.

Our analysis reveals two significant differences between the  Brightest Flare in the current work and typical X-ray flares: (1) the  Brightest Flare exhibits a distinct $L_{\text{peak}}$--Duration and $L_{\text{iso}}$--Duration relation that significantly deviates from typical X-ray flares; (2) the temporal and spectral correlations in the  Brightest Flare show no significant deviation from those observed in typical GRBs.

\begin{deluxetable}{lll}[H]
\tabletypesize{\tiny}
\renewcommand{\arraystretch}{1.2} 
\tablewidth{1pt}
 \setlength{\tabcolsep}{4pt}
\tablecaption{Comparison of the parameters of the  Brightest Flare with those of typical X-ray flares \label{table2}}
\tablehead{
\colhead{Parameter} & \colhead{Brightest Flare} & \colhead{Typical Flare}
}
\startdata
$\bm{\rm Peak~Flux}$ ($\rm erg\cdot cm^{-2}\cdot s^{-1}$) & $(7.30 \pm 0.18) \times 10^{-4}$ & $10^{-14} \sim 10^{-8}$ \\
$\bm{\rm Fluence}$ ($\rm erg\cdot cm^{-2}$) & $(3.49 \pm 0.05) \times 10^{-3}$ & $10^{-9} \sim 10^{-6}$ \\
$\bm{\rm Fluence\ (Brightest \ pulse)}$ ($\rm erg\cdot cm^{-2}$) & $(2.75 \pm 0.04) \times 10^{-3}$ & $10^{-9} \sim 10^{-6}$ \\
$\bm{L_{\text{peak}}}$ ($\rm erg\cdot s^{-1}$) & $(4.51 \pm 0.12) \times 10^{52}$ & $10^{43} \sim 10^{51}$ \\
$\bm{L_{\text{iso}}}$ ($\rm erg\cdot s^{-1}$) & $(1.27 \pm 0.17) \times 10^{52}$ & $10^{43} \sim 10^{50}$ \\
$\bm{L_{\text{iso}}\rm \ (Brightest \ pulse)}$ ($\rm erg\cdot s^{-1}$) & $(1.80 \pm 0.25) \times 10^{52}$ & $10^{43} \sim 10^{50}$ \\
$\bm{E_{\text{iso}}}$ ($\rm erg$) & $(1.82 \pm 0.24) \times 10^{53}$ & $10^{48} \sim 10^{52}$ \\
$\bm{E_{\text{iso}}\rm \ (Brightest \ pulse)}$ ($\rm erg$) & $(1.44 \pm 0.20) \times 10^{53}$ & $10^{48} \sim 10^{52}$ \\
$\bm{T_{\text{rise}}}$ (s) & $0.80 \pm 0.01$ & $10^{-1} \sim 10^{4}$ \\
$\bm{T_{\text{decay}}}$ (s) & $1.79 \pm 0.02$ & $1 \sim 10^{5}$ \\
$\bm{T_{\text{peak}}}$ (s) & $295.47 \pm 0.02$ & $10^{2} \sim 10^{6}$ \\
$\bm{\rm Duration}$ (s) & $16.40 \pm 0.01$ & $1 \sim 10^{6}$ \\
$\bm{\rm Duration\ (Brightest \ pulse)}$ (s) & $9.20 \pm 0.12$ & $1 \sim 10^{6}$ \\
\enddata
\end{deluxetable}

The first difference indicates that, while the duration of the  Brightest Flare does not differ significantly from that of typical X-ray flares, their exceptionally high values $L_{\text{peak}}$, $L_{\text{iso}}$, and $E_{\text{iso}}$ result in a highly deviant statistical behavior related to duration.

The second difference indicates that the correlation between the $L_{\text{peak}}$--$L_{\text{iso}}$ and $L_{\text{peak}}$--$E_{\text{iso}}$ energy relationships, as well as the correlation between the $T_{\text{decay}}$--$T_{\text{rise}}$ and Duration--$T_{\text{peak}}$ time scales suggest that compared to typical X-ray flares, although the  Brightest Flare exhibits the highest luminosity and energy, they still follow the same statistical relationships as typical X-ray flares within the error range.

\subsection{Origin of this BOAT flare}\label{3.4}

In our work, we present more precise observational evidence supporting the hypothesis that flares originate from late activities of the central engine. In previous studies of GRB 221009A \citep{1_09A_Gecam}, the jet opening angle was measured to be only $0.7^\circ$. This narrow angle probably implies the existence of a special central engine that can produce a highly collimated jet. A plausible explanation is that a highly magnetized central engine may generate such a jet with a specific angular structure through the Blandford-Znajek mechanism \citep{1_09A_Gecam}.

\cite{E1_BOAT_flare} proposed that the burst process of GRB 221009A can be explained by a unique two-component jet model. This model consists of a Poynting-flux-dominated jet composition and a broader angular structured jet wing dominated by matter. The narrow jet has an extremely small opening angle ($\theta \sim 0.6^\circ$). Non-thermal radiation dominates in the MeV energy band. Furthermore, the narrow jet exhibits a high Lorentz factor ($\Gamma \sim 600$). The Brightest Flare observed in GRB 221009A occurred from $T_0$+500\,s to $T_0$+520\,s and its extreme brightness may be due to its very narrow jet \citep{E1_BOAT_flare}., The non-thermal radiation components of the BOAT flare in this study further support this hypothesis. Although such narrow jets may be common in GRBs, they may not have been detected in previous observations because of their specific directionality or observational limitations.

Because there is a period dominated by the afterglow component prior to the BOAT flare, we consider the following physical scenario to be likely: during the GRB 221009A outburst the central engine first underwent violent activity that released the bulk of the energy (main emission), then became quiescent as the afterglow phase began, and later reactivated to produce the flare.

\cite{A114_09A_syn} demonstrated that the spectrum of the flare is consistent with a synchrotron radiation model incorporating a decaying magnetic field with GECAM-C data. They further revealed that the physical mechanism of the flare observed during $T_0$+500\,s $\sim$ $T_0$+520\,s in GRB 221009A is consistent with that of the main emission. In the scenario of electron acceleration during the flare phase, the evolution of the electron energy distribution $N(\gamma, t^{\prime} )$ remains governed by the electron continuity equation (Eq.~\ref{eq_flare_e}).

\begin{equation}
\frac{\partial}{\partial t^{\prime}}\left(\frac{dN_{\text{e}}}{d\gamma_{\text{e}}}\right)+\frac{\partial}{\partial\gamma_{\text{e}}}\left[\dot{\gamma}_{\text{e}}\left(\frac{dN_{\text{e}}}{d\gamma_{\text{e}}}\right)\right]=Q(\gamma_{\text{e}},t^{\prime})
\label{eq_flare_e}
\end{equation}

where ${Q(\gamma_{\text{e}},t^{\prime})}$ represents the new injection rate of relativistic electrons. The fitting results indicate that electrons are still predominantly cooled by synchrotron radiation.

The fitting results of the flare by \cite{A114_09A_syn} support the magnetically dominated nature of the jet. Both the flare and the prompt emission exhibit similar evolutionary trends in their physical parameters, indicating that synchrotron radiation dominates their emission mechanisms.

\section{Summary}

In this study, we conducted a comprehensive analysis of the spectral and temporal properties of the flare phase (composed of multiple flares) of the BOAT GRB 221009A, spanning from $T_0$+350\,s to $T_0$+600\,s. Before conducting a detailed analysis, we have demonstrated that the BOAT flare is indeed a flare based on temporal structure of flare and fluence ratio between flare and prompt emission. Particular attention was paid to the  Brightest Flare (i.e., the main phase of the BOAT flare), which occurred from $T_0$ + 500\,s to $T_0$ + 520\,s, using high-resolution data from GECAM-C.

In previous observations of GRB flares by \textit{Swift}, \textit{Fermi}/GBM, and other detectors, the structural characteristics of typical X-ray flares have typically not been detected in the MeV energy band \citep{A3_2007_flare_statis1, A4_2007_flare_statis2, A14_2015_Mev_GeV,A20_2024_gamma_ray_flare}. The high temporal resolution with high statistics structures of flares have also never been detected in the MeV band. The  Brightest Flare has, for the first time, clearly revealed the light curve structure of a GRB flare in the MeV band, as shown in Figure~\ref{fig_lc}. The peak count rate of this flare reaches approximately 40,000 counts/s. The rapid variability and the multipulse superimposed structure of the light curve provide direct observational evidence that the flare and the prompt emission originate from the same physical mechanism.

The $E_{\text{peak}}$ of the  Brightest Flare consistently exceeds 100\,keV. A comprehensive spectral fit to the  Brightest Flare shows that $E_{\text{peak}}$ is about 300 keV. Previous studies have shown that some GRB flares can be fitted with the Band function and CPL model in their spectrum, and their $E_{\text{peak}}$ were mainly concentrated around 1\,keV, rarely exceeding 100\,keV \citep{A3_2007_flare_statis1,A4_2007_flare_statis2,A14_2015_Mev_GeV}. The Brightest Flare has the highest $E_{\text{peak}}$ flare observed to date (Figure~\ref{fig_ratio_Ep} Panel (b)). The complete light curve structure in the $\gamma$-ray and MeV bands, along with the persistent presence of $E_{\text{peak}}$ within the $\gamma$-ray band throughout the flare, collectively confirm that the  Brightest Flare is a rarely detected $\gamma$-ray flare. We note that there may be a population of GRBs with gamma-ray flares which are just too weak to be clearly detected and verified. In addition, the  Brightest Flare exhibits a record breaking $E_{\text{iso}}$ = $1.82 \times 10^{53} \, \text{erg}$. We find that the  Brightest Flare and typical X-ray flares share similar statistical properties. The exceptional brightness of the  Brightest Flare may be attributed to its highly magnetized, narrow jet structure, which aligns with its non-thermal radiation predominantly observed in the keV to MeV energy band.

Finally, we stress that this is the first observation of a high temporal resolution with high statistics light curve of the GRB flare in the keV-MeV band, bridging the last gap between prompt emission and flare.
We provide strong observational evidence that the flare and the prompt emission have the same physical mechanism. In particular, the rapidly varying multipulse superimposed light curve structure observed in the energy band similar to that of prompt emission (6\,keV - 6\,MeV) offers direct observational support for this view. In simple terms, by analyzing the BOAT flare with methods and precision identical to those applied to the prompt emission, we obtained similar observational results. 
Therefore, our detailed analysis of the BOAT flare improves our understanding of the physical processes involved in flares of GRBs and this flare bridges the last gap between prompt emission and flare in GRB

\begin{acknowledgments}
This work is supported by  
the National Natural Science Foundation of China (Grant No. 12494572, 12273042), the Strategic Priority Research Program of the Chinese Academy of Sciences (Grant No. XDA30050000,
XDB0550300
)
and the National Key R\&D Program of China (2021YFA0718500).
The GECAM (Huairou-1) mission is supported by the Strategic Priority Research Program on Space Science (Grant No. XDA15360000) of Chinese Academy of Sciences. We appreciate the GECAM team and the SATech-01 satellite team who made this observation possible.
\end{acknowledgments}

\clearpage
\appendix

\begin{figure}[H]
\centering
\begin{tabular}{cc}
  \begin{minipage}{0.49\textwidth}
      \includegraphics[width = \textwidth]{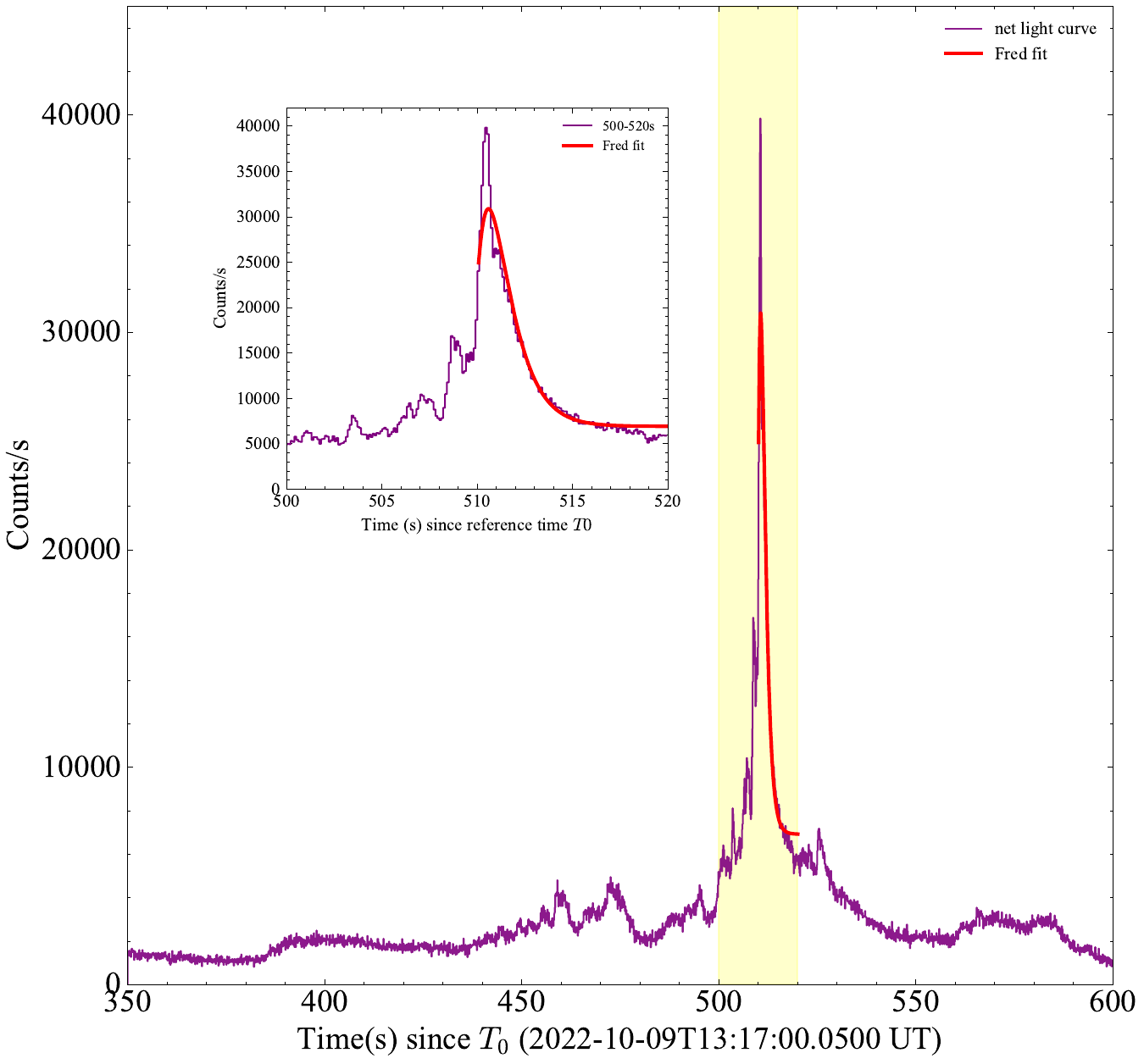}
\subcaption{}
   \end{minipage}
    \begin{minipage}{0.47\textwidth}
        \includegraphics[width = \textwidth]{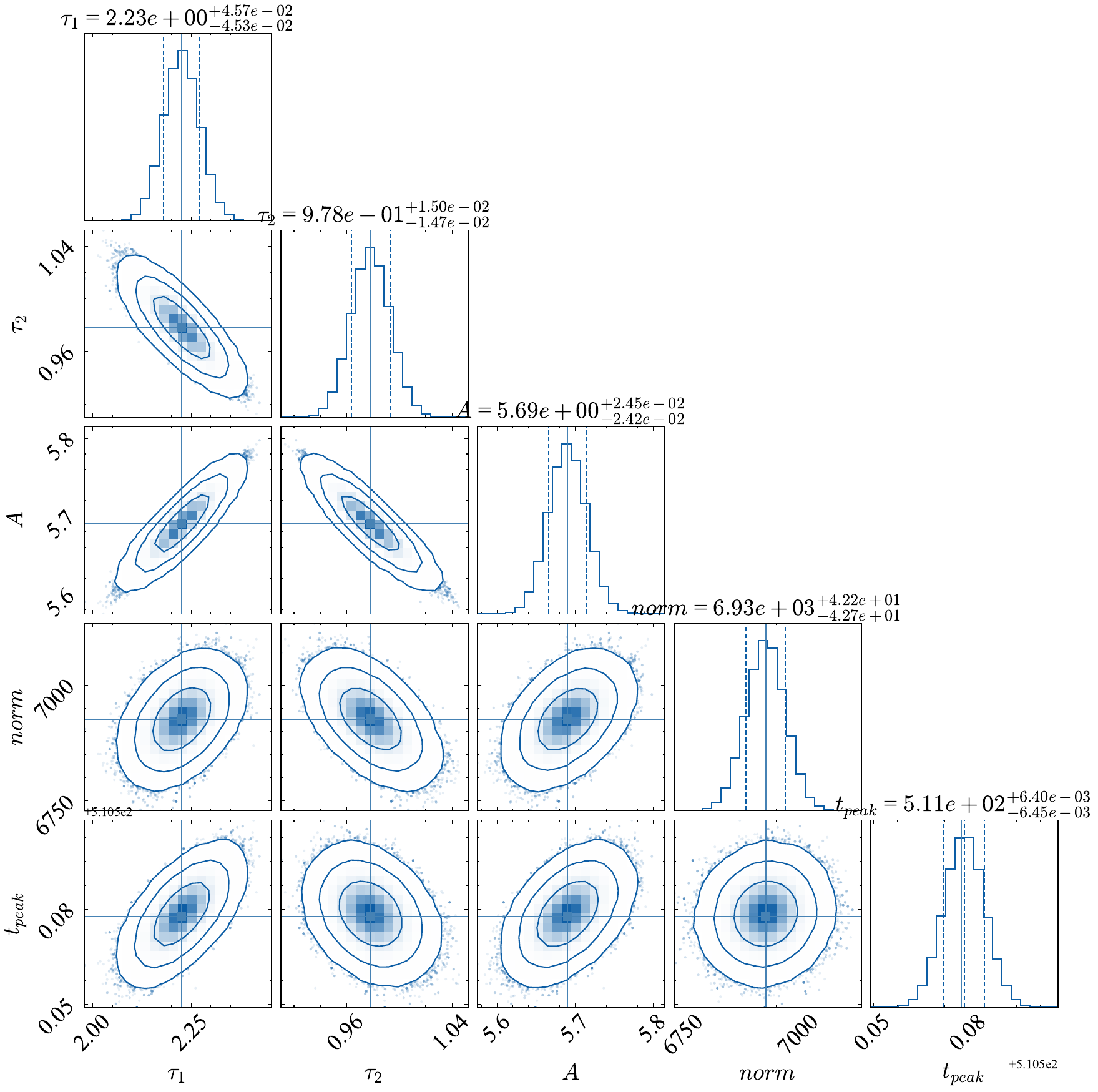}
\subcaption{}
    \end{minipage}
\end{tabular}
  \caption{\label{figA1}\small \textbf{Panel (a):} The red curve shows the Norris05 function fit to the peak of the brightest flare, while the purple curve represents the net light curve. The  Brightest Flare is highlighted in yellow in the main panel. A detailed view of the net light curve and its fit for the main burst is provided in the left inset. \textbf{Panel (b):} The corner plot of the Norris05 function. }

\end{figure}

\begin{table}[H]
\begin{center}
\caption{Flux lightcurve fitting result of GRB 221009A.}\label{table_fit}
\begin{tabular*}{\hsize}{@{}@{\extracolsep{\fill}}cccccccccc@{}}
\toprule
Phase & model & $\alpha_1$ & $\alpha_2$ & $t_{{\rm b}}$ & $\log_{10}A$ & $\chi^2$/dof\\ 
 &  &  &  & (s) & (erg$\cdot$cm$^{-2}\cdot$s$^{-1}$) & \\ 
\midrule
${\rm Flare_1}$ & ${\rm SBPL_1}$ & -8.02$^{+0.06}_{-0.03}$ & 2.83$^{+0.04}_{-0.06}$ & 107.34$^{+1.12}_{-3.01}$ & -5.22$^{+0.02}_{-0.01}$ & \multirow{6}{*}{632/36}\\
${\rm Flare_2}$ & ${\rm SBPL_2}$ & -7.95$^{+0.07}_{-0.09}$ & 3.52$^{+0.08}_{-0.05}$ & 173.23$^{+2.01}_{-1.42}$ & -5.13$^{+0.04}_{-0.03}$ &\\
${\rm Flare_3}$ & ${\rm SBPL_3}$ & -7.12$^{+0.12}_{-0.21}$ & 16.14$^{+0.09}_{-0.13}$ & 247.36$^{+0.71}_{-0.82}$ & -4.82$^{+0.01}_{-0.02}$ &\\
${\rm Flare_4}$ & ${\rm SBPL_4}$ & -93.15$^{+0.56}_{-0.62}$ & 75.42$^{+0.63}_{-0.56}$ & 285.56$^{+1.57}_{-1.75}$ & -3.41$^{+0.05}_{-0.06}$ &\\
${\rm Flare_5} $& ${\rm SBPL_5}$ & -20.21$^{+0.11}_{-0.09}$ & 16.12$^{+0.07}_{-0.07}$ & 347.23$^{+2.03}_{-1.52}$ & -4.85$^{+0.02}_{-0.03}$ &\\
${\rm Afterglow}$ & BPL & 1.04$^{+0.01}_{-0.01}$ & 1.82$^{+0.06}_{-0.06}$ & 1020.58$^{+27.00}_{-26.12}$ & -6.46$^{+0.02}_{-0.02}$ & \\
\botrule
\end{tabular*}
\end{center}
\end{table}

\begin{longrotatetable}
\centering
\begin{deluxetable*}{@{}cccccccccccc@{}}
\tabletypesize{\tiny}
\renewcommand{\arraystretch}{0.2}
\setlength{\tabcolsep}{8pt} 
\tablecaption{Time resolved spectra fitting result of the GRB 221009A flare}
\label{table_spec}
\tablehead{
Type & Time Interval & Model & $\alpha$ & $\beta$ & $E_{\rm peak}$ & $Index$ & kT & Flux & cstat/dof & BIC & Best model\\
 & (s) &  &  &  & (keV) &  & (keV) & $10^{-5} \text{erg} \cdot \text{cm}^{-2} \cdot \text{s}^{-1} $ &  &  &
}
\renewcommand{\arraystretch}{0.9}
\startdata
 \multirow{21}{*}{\textbf{S-1}} & \multirow{7}{*}{[350.0-500.0]} & Band & -1.59$^{+0.01}_{-0.01}$ & -2.31$^{+0.04}_{-0.05}$ & 137.52$^{+4.55}_{-4.39}$ & - & - & 1.79$^{+0.04}_{-0.04}$ & 317.65/304 & 346.0 & \multirow{7}{*}{Band} \\
 &  & CPL & -1.62$^{+0.01}_{-0.01}$ & - & 156.34$^{+4.67}_{-3.91}$ & - & - & 1.48$^{+0.02}_{-0.02}$ & 388.03/305 & 410.0 &  \\
 &  & PL & - & - & - & 1.85$^{+0.00}_{-0.00}$ & - & 2.93$^{+0.02}_{-0.02}$ & 1286.59/306 & 1303.0 &  \\
 &  & Band+PL & -1.60$^{+0.01}_{-0.01}$ & -2.32$^{+0.05}_{-0.06}$ & 139.40$^{+4.22}_{-4.03}$ & 6.97$^{+2.43}_{-4.52}$ & - & 1.79$^{+0.05}_{-0.05}$ & 318.29/302 & 358.0 &  \\
 &  & CPL+PL & -1.62$^{+0.01}_{-0.02}$ & - & 151.30$^{+6.96}_{-8.02}$ & 3.66$^{+5.63}_{-2.79}$ & - & 1.50$^{+0.30}_{-0.04}$ & 389.60/303 & 424.0 &  \\
 &  & Band+BB & \multicolumn{8}{c}{Unconstrained} &  \\
 &  & CPL+BB & -1.62$^{+0.01}_{-0.02}$ & - & 155.10$^{+4.59}_{-3.17}$ & - & 0.77$^{+4.19}_{-0.55}$ & 1.48$^{+0.19}_{-0.03}$ & 387.59/303 & 421.0 &  \\
 \cline{2-11}
 & \multirow{7}{*}{[500.0-520.0]} & Band & -1.29$^{+0.01}_{-0.01}$ & -1.88$^{+0.10}_{-0.13}$ & 300.91$^{+31.21}_{-31.40}$ & - & - & 17.43$^{+0.23}_{-0.24}$ & 615.95/304 & 644.0 & \multirow{7}{*}{Band} \\
\enddata
\tablecomments{ All time intervals are relative to the trigger time $T_0$. 
Uncertainties represent 1$\sigma$ confidence intervals.}
\end{deluxetable*}
\end{longrotatetable}

\clearpage
\bibliography{sample7}{}
\bibliographystyle{aasjournalv7}

\end{document}